\documentclass[showpacs,prl,onecolumn,aps,superscriptaddress,preprintnumbers,nofootinbib]{revtex4}
\usepackage[T1]{fontenc}
\usepackage[latin9]{inputenc}
\usepackage{graphicx,hyperref}

\usepackage{amsmath,amssymb}
\usepackage{epsfig}
\usepackage{graphicx}
\usepackage{amsmath}
\usepackage{amsfonts}
\usepackage{mathrsfs}
\usepackage{pifont}
\usepackage{relsize}
\usepackage{mathtools}
\def\slashchar#1{\setbox0=\hbox{$#1$}     		
   \dimen0=\wd0                                 	
   \setbox1=\hbox{/} \dimen1=\wd1               	
   \ifdim\dimen0>\dimen1                        	
      \rlap{\hbox to \dimen0{\hfil/\hfil}}      	
      #1                                        	
   \else                                        	
      \rlap{\hbox to \dimen1{\hfil$#1$\hfil}}   	
      /                                         	
   \fi}

\renewcommand{\vec}{\boldsymbol}
\newcommand{\beq}{\begin{equation}}
\newcommand{\eeq}{\end{equation}}
\newcommand{\bea}{\begin{eqnarray}}
\newcommand{\eea}{\end{eqnarray}}
\newcommand{\baa}{\begin{array}}
\newcommand{\eaa}{\end{array}}

\def\eq#1{{Eq.~(\ref{#1})}}

\def\fig#1{{Fig.~\ref{#1}}}
\newcommand{\intl}{\int\limits}

\newcommand{\bas}{\bar{\alpha}_S}

\newcommand{\nn}{\nonumber}

\newcommand{\h}{\frac{1}{2}}

\newcommand{\x}{\vec{x}}
\newcommand{\vb}{\vec{b}}
\newcommand{\z}{\vec{z}}

\newcommand{\al}{\alpha}

\newcommand{\la}{\lambda}

\newcommand{\Lb}{\left(}
\newcommand{\Rb}{\right)}

\newcommand{\dy}{\delta Y}
\newcommand{\pp}{\partial}

\renewcommand{\vec}[1]{\boldsymbol{#1}}

\newcommand{\ga}{\gamma}
\newcommand{\dY}{\delta \tilde{Y}}
\newcommand{\zz}{\tilde{z}}
\newcommand{\Y}{\tilde{Y}}
\newcommand{\xt}{\tilde{\xi}}
\begin{document}
\title{ Homotopy solution to non-linear evolution for heavy nuclei}
\author{Carlos Contreras}
\email{carlos.contreras@usm.cl}
\affiliation{Departamento de F\'isica, Universidad T\'ecnica Federico Santa Mar\'ia,  Avda. Espa\~na 1680, Casilla 110-V, Valpara\'iso, Chile}
\author{ Eugene ~ Levin}
\email{leving@tauex.tau.ac.il, eugeny.levin@usm.cl}
\affiliation{Departamento de F\'isica, Universidad T\'ecnica Federico Santa Mar\'ia,  Avda. Espa\~na 1680, Casilla 110-V, Valpara\'iso, Chile}
\affiliation{Centro Cient\'ifico-
Tecnol\'ogico de Valpara\'iso, Avda. Espa\~na 1680, Casilla 110-V, Valpara\'iso, Chile}
\affiliation{Department of Particle Physics, School of Physics and Astronomy,
Raymond and Beverly Sackler
 Faculty of Exact Science, Tel Aviv University, Tel Aviv, 69978, Israel}
\author{Rodrigo Meneses}
\email{rodrigo.meneses@uv.cl}
\affiliation{Escuela de Ingenier\'\i a Civil, Facultad de Ingenier\'\i a, Universidad de Valpara\'\i so,Avda  General Cruz 222 , Valpara\'\i so, Chile}

\date{\today}


\begin{abstract}

In the paper we suggest the homotopy method for solving of the non linear evolution equation. This method consists of two steps. First is  the analytical solution for the linearized version of the non-linear evolution deep in the saturation region. Second, the perturbative procedure is suggested to take into account the remaining parts of the non-linear corrections. It turns out that these corrections are rather small and can be estimated in the regular  iterative procedure.

 \end{abstract}
\maketitle

\vspace{-0.5cm}
\tableofcontents





\section{Introduction}

The Colour Glass Condensate approach\cite{GLR,MUQI,MUCD,MV,KLBOOK,BK,JIMWLK}(CGC) 
 has reached a mature stage  and has become the common language to discuss high energy scattering where the dense system of partons (quarks and gluons) is produced.
 The most theoretical progress has been reached in the description of  dilute-dense parton system  scattering\cite{REV}.  The deep inelastic scattering (DIS) of electron with hadrons  is well known example of such process. For these processes we know their main qualitative features as well as 
the non-linear evolution equation\cite{BK,JIMWLK}, that describes them. For DIS with  protons it has been shown that at high energies the density of partons reaches the saturation and dipole-hadron amplitude tends to approach unity\cite{GLR}. The analytical form of the scattering amplitude at high energies has been found\cite{LT}. At high energies the scattering amplitude shows the scaling behaviour \cite{LT,BALE,SGBK,IIM}, which leads to the  amplitude being  a function of one variable $x^2_{10} Q^2_s(Y, b)$\footnote{$x_{10}$ is the size of scattering dipole  and $Q_s$ is a new  momentum scale (saturation momentum). Generally speaking each dipole has   coordinates $x_i, y_i$, where $x_i$ and $y_i$ are coordinates of quark and antiquark, respectively. We introduce $r_i= x_i - y_i$ for the size of the dipole and $b_i = \h(x_i + y_i).$ for the impact parameter. However, we often use the  alternative notations: $x_{ik} = | x_i - y_{k = i+1}|$, for the size of the dipole.
 }. In two  regions we have the analytical solution to the non-linear Balitsky-Kovchegov\cite{BK} (BK)  equation: (i) for $x^2_{10} Q^2_s(Y,b)\,\,\gg\,\,1$\cite{LT}  and (ii) for  $x^2_{10} Q^2_s(Y,b)\,\,\sim\,\,1$\cite{MUT}.  
 
 In other words, the non-linear evolution BK equation  for the scattering amplitude     $N_{ik}=N\Lb Y, \x_{ik},\vb\Rb$ has the following general form\footnote{We write the BK equation for large impact parameters $b\,\gg\,r_{i k}$.}:
  \beq \label{BK}
 \frac{\partial N_{01}}{\partial Y}\,=\,\bas\int \frac{d^2\,x_{02}}{2 \pi} \frac{ x^2_{01}}{x^2_{02}\,x^2_{12}}\Big\{ N_{02} + N_{12} - N_{02}N_{12} - N_{01}\Big\}
 \eeq
  and it  has the analytical solutions in these two regions. Indeed, for  $x^2_{10} Q^2_s(Y)\,\,\gg\,\,1$  the scattering amplitude approaches unity and has the form\cite{LT}:
  \beq \label{BKLT1}
N_{01}\Lb z\Rb\,\,\,=\,\,1\,\,\,-\,\,\,\mbox{Const}\,\exp\Big(  - \frac{z^2}{2\,\kappa}\Big)
 \eeq 
 where $z$ and $\kappa$ are determined by the following equations\footnote{$\chi\Lb \gamma\Rb$ is the BFKL kernel\cite{BFKL,LIP} in anomalous dimension ($\gamma$) representation.}:
 \beq \label{GACR}
\kappa \,\,\equiv\,\, \frac{\chi\Lb \gamma_{cr}\Rb}{1 - \gamma_{cr}}\,\,=\,\, - \frac{d \chi\Lb \gamma_{cr}\Rb}{d \gamma_{cr}}~~~\,\,\,\mbox{and}\,\,\,~~~\chi\Lb \gamma\Rb\,=\,\,2\,\psi\Lb 1 \Rb\,-\,\psi\Lb \gamma\Rb\,-\,\psi\Lb 1 - \gamma\Rb
 \eeq 
  and for 
    new variable $z$ we have 
 \beq \label{z}
 z\,\,=\,\,\ln\Lb x^2_{01}\,Q^2_s\Lb Y, b\Rb\Rb\,\,\, =\,\,\,\,\bas\,\kappa \,\Lb Y\,-\,Y_A\Rb\,\,+\,\,\xi\,\,\,=\,\,\xi_s\,\,+\,\,\xi
 \eeq 
 where $\xi \,\,=\,\,\ln\Lb x^2_{10}\,Q^2_s\Lb Y=Y_A, b \Rb\Rb$ and $Q^2_s(Y, b)$ is equal to
 
 \beq \label{QS}
 Q^2_s\Lb Y, b\Rb\,\,=\,\,Q^2_s\Lb Y=Y_A, b\Rb \,e^{\bas\,\kappa \Lb Y - Y_A\Rb}\
 \eeq 
 In \eq{z} and \eq{QS} $Y_A$ denotes $ Y_A = \ln A^{1/3}$, where $A$ is the number of nucleons in a nucleus.
 
 In the vicinity of the saturation scale $x^2_{10}\,Q^2_s\Lb Y,b\Rb\,\sim\,1$ the scattering amplitude has the form\cite{MUT}:
 \beq \label{VQS}
 N_{01}\Lb z\Rb\,\,\,=\,\,\,\mbox{Const} \Lb x^2_{10}\,Q^2_s\Lb Y,b\Rb\Rb^{\bar \gamma}
 \eeq
 with $\bar \gamma = 1 - \gamma_{cr}$. 
 
 From \eq{VQS}  one can see that at $x_{01}\,Q_s=1 (z=0)$ we have the following boundary conditions for the saturation region:
 \beq \label{BCINT}
  N_{01}\Lb z = 0\Rb \,\,=\,\,N_0;~~~~~~~~\frac{d N_{01}\Lb z\Rb}{d\,z}\Bigg{|}_{z=0}\,\,=\,\,\bar{\ga}\,N_0.
  \eeq

 Finally, for $x^2_{10}\,Q^2_s\Lb Y, b\Rb\,\ll\,1$  the scattering amplitude satisfies the linear BFKL equation and is a function of two variables: $Y$ and $\xi$.
 
 We have defined the saturation region as $ x^2_{10}\,Q^2_s\Lb Y,b\Rb\,>\,1$. However, in Refs.\cite{GOST,BEST} it has been noted that actually for very large $x_{10} $ the non-linear corrections are not important and we have to solve linear BFKL equation. This feature can be seen  directly
 from the eigenfunction of this equation.  Indeed, the eigenfunction
 (the scattering amplitude of two dipoles with sizes $r \equiv  x_{10}$ and $R$) has the
 following form \cite{LIP}
\beq \label{EIGENF}
\phi_\gamma\Lb \vec{r} , \vec{R}, \vec{b}\Rb\,\,\,=\,\,\,\Lb \frac{
 r^2\,R^2}{\Lb \vec{b}  + \h(\vec{r} - \vec{R})\Rb^2\,\Lb \vec{b} 
 -  \h(\vec{r} - \vec{R})\Rb^2}\Rb^\gamma~~~~\mbox{with}\,\,0 \,<\,Re\gamma\,<\,1
 \eeq 

One can see that for $r=x_{10} \,>\,min[R, b]$,  $\phi_\gamma$ starts to be small and the non-linear term in the BK equation 
could be neglected.  In this paper we consider the interaction with nucleus  and the  scattering amplitude of the dipole with a nucleus  for the exchange of the BFKL Pomeron   has the following form:
\beq \label{EIGENF1}
A_{d A} \Lb  \vec{r} , \vec{R}_N, \vec{b}\Rb\,=\,\phi_\gamma\Lb \vec{r} , \vec{R}_N, \vec{c}\Rb\,\,S_A\Lb \vec{b} - \vec{c}\Rb d^2 c\eeq
where $R_N$ is the size of a nucleon, $\vec{c}$ is the impact parameter for dipole-nucleon amplitude and $\vec{b} - \vec{c}$ is the position of the nucleon with respect to the center  of the nucleus. $S_A\Lb \vec{b} - \vec{c}\Rb$ is the number of  nucleons in a nucleus.  Since $c \,\ll \,b$   we can integrate over $\vec{c}$ replacing $\vec{b} - \vec{c} $ by $\vec{b}$ and obtain  the expression:
\bea \label{EIGENF2}
A_{d A} \Lb  \vec{r} , \vec{R}_N, \vec{b}\Rb  &=&\int \phi_\gamma\Lb \vec{r} , \vec{R}_N, \vec{c}\Rb\,\,S_A\Lb \vec{b} - \vec{c}\Rb d^2 c = \int d^2 c \, \phi_\gamma\Lb \vec{r} , \vec{R}_N, \vec{c}\Rb\,\,S_A\Lb \vec{b}\Rb \nn\\
&=&  \Lb \frac{r^2}{R^2_N}\Rb^\gamma \!\!R^2_N S_A\Lb \vec{b}\Rb   =   \Lb r^2\,Q^2_s\Lb Y=Y_A, b\Rb \Rb^\gamma\,\eea
Therefore, in this case we can absorb all dependence on the impact parameter in the $b$  dependence of the saturation scale. In \eq{EIGENF2} we implicitly assume that $ r \leq R_N$. 
 The typical process, that we bear in mind, is the deep inelastic scattering(DIS) with a nucleus at $Q^2\, \geq \,1 GeV^2$ and at small values of $x$.

 For getting more quantitative  information, which is needed for comparison with the experimental data,  the BK equation has to be solved numerically. However,  the CGC approach suffers the principle difficulty: the power-like behaviour of the scattering amplitude  at large impact parameters $b$ \cite{KW1,KW2,KW3,FIIM}, which leads the violation of the Froissart theorem \cite{FROI}. The widely accepted phenomenological way to heal this problem is to introduce the non-perturbative dependence of $Q^2_s\Lb Y=0,b\Rb \propto \exp\Lb - \mu\,b\Rb$ (see Refs.\cite{IIMU,BKL,RESH,CLP} and references therein). One can see that 
 a numerical solution  of the BK equation does not allow us to introduce these corrections since the BK equation does not   have an  explicit dependence on $Q_s$.  The standard way of dealing with the non-perturbative corrections is to change the BFKL kernel  introducing a soft scale. In doing this we violate the conformal   symmetry of the kernel,  which is the basis of our understanding of many features of the solution.   We consider
 the non-perturbative corrections to the saturation scale, which the new and the only dimensional scale in our problem, 
 as more attractive way of introducing the soft scale.  Bearing this in mind 
   we need to have the analytical solution in which we can see explicitly the dependence of the scattering amplitude on $Q_s(Y,b)$.
 
 In this paper we have two goals. First we suggest a procedure to find the solution to the BK equation as a  sequence of iterations, based on homotopy approach \cite{HE1,HE2}. This approach gives us a tool to introduce the saturation momentum in our calculation, starting from the first iteration, and allow us the insert the non-perturbative changes in $b$ dependence of $Q_s(b)$. 
 
 The homotopy method we can use in the case when you equation has a general form:
 \beq \label{HOM1}
\mathscr{L}[u] +  \mathscr{N_{L}}[u]=0
\eeq 
 where the linear part $\mathscr{L}[u] $ is a differential  or integral-differential operator, but non-linear part $
 \mathscr{N_{L}}[u] $ has an arbitrary form. For solution we introduce  the following  equation for the homotopy function $ {\cal H}\Lb p,u\Rb$:
 \beq \label{HOM2}
 {\cal H}\Lb p,u\Rb\,\,=\,\,\mathscr{L}[u_p] \,+ \,  p\, \mathscr{N_{L}}[u_p] \,\,=\,\,0
 \eeq
 
 Solving \eq{HOM2} we reconstruct the function
 \beq \label{HOM3}
 u_p\Lb Y,  \x_{10},  \vb\Rb\,\,=\,\, u_0\Lb Y,  \x_{10},  \vb\Rb\,\,+\,\,p\, u_1\Lb Y,  \x_{10},  \vb\Rb \,+\,p^2\, u_3\Lb Y,  \x_{10},  \vb\Rb \,\,+\,\,\dots
 \eeq
 with $\mathscr{L}[u_0] = 0$. \eq{HOM3}  gives  the solution to the non-linear equation at $ p$=1.  The hope is that several  terms in series of \eq{HOM3} will give a good  approximation to the solution of the non-linear  equation. This method has been applied to the solution of the BK equation\cite{SPS},  but we suggest a quite different  procedure that allow us to see the  geometric scaling behaviour as well as the dependence on the saturation scale in explicit form on each stage of the solution.

\begin{figure}
 	\begin{center}
 	\leavevmode
 		\includegraphics[width=8cm]{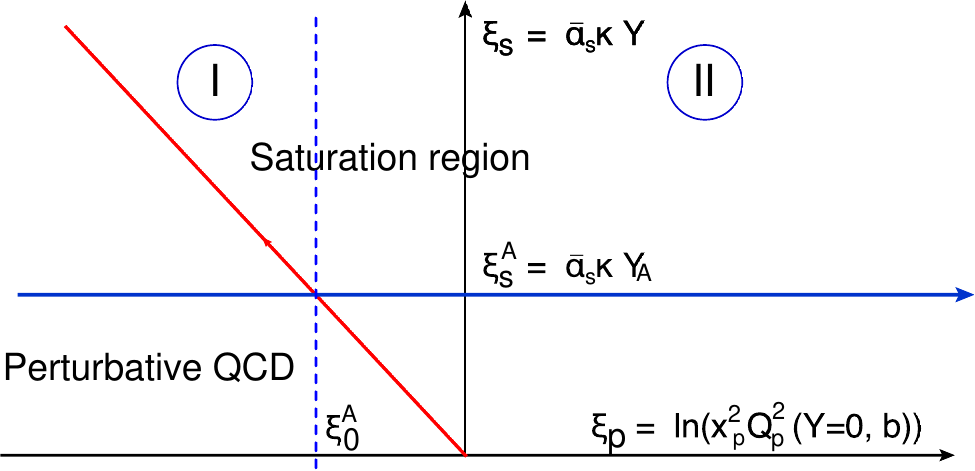}
 	\end{center}
 	\caption{  Saturation region of QCD. The critical line (z=0) is shown in red. The initial condition for scattering with the dilute system of partons (with proton) is given at $\xi_s = 0$. For heavy nuclei the initial conditions are placed at $Y_A = (1/3)\ln \,A \gg\,1$, where $A$ is the number of nucleon in a nucleus. The line, where they are given, is shown in blue.}
 	\label{sat}
 \end{figure}

 The second goal of this paper is to find the efficient procedure to solve the BK equation for interaction with nuclei. In this case, as the initial conditions at $Y=Y_A$  for DIS with nuclei we have McLerran-Venugopalan formula\cite{MV} for the imaginary part of the dipole-nucleus amplitude, which takes the following form\footnote{For the exact form of \eq{MVF} see Ref.\cite{MV},  we took  the simplified version of it from Ref.\cite{GBW}.}  (see \fig{sat})
\beq \label{MVF}
N\Lb x^2_{10}, Y = Y_A, b \Rb\,\,\,=\,\,1\,\,\,-\,\,\,\exp\Big( - r^2\,Q^2_s\Lb Y=Y_A, b\Rb/4\Big)\,\,=\,\,1\,\,\,-\,\,\exp\Lb -\frac{1}{4}\, e^\xi\Rb
\eeq
where $r = x_{10}$ and we will use this notation for the size of the dipole below.

 One can see from \fig{sat}, that for large $Y_A\,=\,\frac{1}{3} \ln A$, these initial conditions, which violate the geometric scaling behaviour, give the scattering amplitude in the saturation region.
Comparing \eq{MVF} and \eq{BKLT1} we see that the geometric scaling behaviour  cannot be correct in the entire saturation region. This problem is not new and we have discussed it in our previous papers \cite{LTHI,KLT,CLM}. Having a new iteration procedure we hope to suggest a practical way to take  into account the initial conditions of \eq{MVF}.
 
 \eq{MVF} in the region, where we can trust perturbative QCD, has the form: $\frac{1}{4}e^{\xi_0^A}$.  Hence the equation for $\xi_0^A$(see \fig{sat}) takes the following form: 
 \beq \label{XI0}
 \frac{1}{4}e^{\xi_0^A}\,=\,\,\mbox{Const} \Lb x^2_{10}\,Q^2_s\Lb Y,b\Rb\Rb^{\bar \gamma}
 \eeq

 It should be noted that \eq{BCINT}  for any value of $z$ in the vicinity of the saturation scale:  $ x^2_{10}\,Q^2_s\Lb Y, b \Rb\,\sim\,1 $.

In the next section we introduce the non-linear Balitsky-Kovchegov equation in the momentum space and discuss the homotopy approach for solving this equation. The
suggested approach consists of two steps. First, we introduce the linearized  vesion of the non-linear equation deep in the saturation region.Using the solution of this linear equation we develop the homotopy approach that   gives us a regular procedure how to take into account the non-linear correction which could be sizable in the saturation region. In section 3 we discuss in detail the first two iterations of the homotopy approach. In particular, we demonstrate that the first iteration gives a rather small contribution. This, in our opinion, indicates that the homotopy approach gives the regular way to account the non-linear corrections, which are non-perturbative in their origin, in the perturbative way after subtracting the contributions of them  deep  in the saturation region. In conclusions we summarize our results.

\section{Equation and solution in the momentum space}
\subsection{Rewriting BK equation in the convenient for the homotopy approach form}
We re-write the Balitsky-Kovchegov equation of \eq{BK}  in the momentum space introducing
\beq \label{MR}
N\Lb r^2, b, Y\Rb\,\,=\,\,r^2 \,\int \frac{d^2 k_\perp}{ 2 \pi}\,e^{ i \vec{k}_\perp \cdot \vec{r}}\,\widetilde{N}\Lb k_\perp, b, Y\Rb \,=\,r^2 \intl^\infty_0 k_\perp d k_\perp J_0\Lb r k_\perp\Rb \,\widetilde{N}\Lb k_\perp, b, Y\Rb\eeq
In \eq{MR} we assume that the amplitude has  radial symmetry and it takes the form\cite{GLR,KOV}
\beq \label{BKMR}
\frac{\partial \widetilde{N}\Lb k_\perp, b, Y\Rb}{\partial Y}\,\,=\,\,\bas \Bigg\{ \chi\Lb -\,\frac{\partial}{\partial \tilde{\xi}}\Rb
\widetilde{N}\Lb k_\perp, b, Y\Rb\,\,\,-\,\, \widetilde{N}^2\Lb k_\perp, b, Y\Rb\Bigg\}
\eeq
where
\beq \label{XIT}
\tilde{\xi}\,\,=\,\,\,\,\ln\Lb Q^2_s\Lb Y = Y_A, b\Rb/ k^2_\perp\Rb~~~\mbox{and}~~~\tilde{z}\,\,=\,\,\bas \kappa \Lb Y - Y_A\Rb \,\,+\,\,\tilde{\xi}\,\,=\,\,\ln \Lb Q^2_s\Lb Y, b\Rb/k^2_\perp\Rb
\eeq

The advantage of the non-linear equation in \eq{BKMR}, that the non-linear term depends only on external variables  and does not contain the integration over momenta.  The  BFKL kernel\cite{BFKL}: $  \chi\Lb -\,\frac{\partial}{\partial \tilde{\xi}}\Rb$, 
can be written as the series over positive powers of $\partial/\partial \tilde{\xi}$ except of the first term

\beq \label{FIT}
\frac{1}{\gamma}\, \widetilde{N}\Lb\ga, b, Y\Rb\,\,\, \to\,\,\, \intl^{k^2_\perp}\frac{ d k'^2_\perp}{k'^2_\perp}  \widetilde{N}\Lb k'_\perp, b, Y\Rb
\eeq
 where $\widetilde{N}\Lb\ga, b, Y\Rb$ is the image of the scattering amplitude in $\gamma$-representation:
\beq \label{NGR} 
\widetilde{N}\Lb k_\perp, b, Y\Rb= \intl^{\epsilon + i \infty}_{\epsilon - i \infty}\!\!\! \frac{d \gamma}{2 \pi i} \Lb k^2_\perp\Rb^\gamma\widetilde{N}\Lb\ga, b, Y\Rb.
\eeq

Differentiating \eq{BKMR} over $\tilde{\xi}$ one can see that  it can be re-written as
\bea \label{BKMR1}
&&\frac{\partial^2 \widetilde{N}\Lb k_\perp, b, Y\Rb}{\partial Y\,\partial \xt} \,\,\,=\\
&& \,\,\,\bas\Bigg\{ \chi_{0}\Lb-\frac{\partial}{\partial \tilde{\xi}}   \Rb\,\, \frac{\partial \widetilde{N}\Lb k_\perp, b, Y\Rb}{\partial\,\xt}\,\,+\,\,  \widetilde{N}\Lb k_\perp, b, Y\Rb\,\,-\,\,2 \frac{\partial \widetilde{N}\Lb k_\perp, b, Y\Rb}{\partial\,\xt}  \widetilde{N}\Lb k_\perp, b, Y\Rb \Bigg\}\nn
\eea
with $\chi_{0}(\gamma)\,=\, \chi(\gamma)\,\,-\dfrac{1}{\gamma}$ where $\chi\Lb \gamma\Rb$   is given by \eq{GACR}.

Introducing the variable $\tilde{z}$ instead of $\tilde{\xi}$ and the new functions $M$ and $\Phi$ as
\beq \label{M}
\frac{\partial \widetilde{N}\Lb k_\perp, b, Y\Rb}{\partial\,\zz}\,\,=\,\,\h +M\Lb\tilde{z}, b, \dY\Rb ~~~~\mbox{or}~~~~\widetilde{N}\Lb  \tilde{z},b, \dY\Rb\,\,=\,\,\h \tilde{z}\,\,+\,\,\int\limits^{\tilde{z}}_{0} d \tilde{z}'  M\Lb\tilde{z}',b,Y\Rb\,\,+\,\,\Phi\Lb \dY\Rb
\eeq
we can re-write \eq{BKMR1} in the form
\beq \label{BKMR2}
\frac{\partial M\Lb\tilde{z}, b, \dY\Rb}{\partial \dY}\,\,+\,\, \kappa\,\frac{\partial M\Lb\tilde{z}, b, \dY\Rb}{\partial \tilde{z}} \,\,\,=
 \chi_{0}\Lb \frac{\partial}{\partial \tilde{z}}\Rb\ M\Lb \tilde{z},  b, \dY\Rb \,\,-\,\,\,\tilde{z}\,M\Lb \tilde{z}, b, \dY\Rb\,\ -2\,\Phi\Lb \dY\Rb\,M\Lb \zz,\dY\Rb\,\,-\dfrac{\partial A}{\partial \zz}
\eeq
with $\dY = \overline{\alpha}_{S} \Lb Y  -  Y_A\Rb$ and
\beq \label{AUXFUN}
A(\zz,b,\dY)=\left( \int_{0}^{\zz}M(z',b,\dY)dz' \right)^{2}
\eeq
 In \eq{M} we introduce a new function $M$, which  approaches zero at large $\zz$, since $\h$ is the correct asymptotic behaviour in the momentum representation. It corresponds  to the  amplitude in the coordinate representation, that is equal to unity deep in the saturation region.

Looking for  $N(r,b,Y)$, that depend only on |r|, we can rewrite \eq{MR} as follows
\beq\label{HANKEL:RAD}
\widetilde{N}(k_{\perp},b,\dY)=\intl_{0}^{\infty}rdr\dfrac{J_{0}(rk_{\perp})}{r^{2}}N(r,b, \dY)
\eeq 
On the other hands, differentiating \eq{HANKEL:RAD} with respect to $k_{\perp}$ we have
\beq\label{DIFHANKEL:RAD}
\dfrac{\pp \widetilde{N}(k_{\perp},b, \dY)}{\pp k_{\perp}}=-\intl_{0}^{\infty}drJ_{1}(rk_{\perp})N(r,b,\dY) 
\eeq 
From \eq{M} and \eq{DIFHANKEL:RAD} we obtain for $M(\zz,b,\dY)$ the following formula:
\beq \label{M1}
M(\zz,b, \dY)\,\,=\,\,k_\perp\intl_{0}^{\infty}drJ_{1}(rk_{\perp})\Bigg\{
\frac{N(r,b,\dY)\,\,-\,\,1}{2}\Bigg\} 
\eeq

\subsection{Homotopy approach}
We  rewrite the non-linear evolution equation in momentum representation (see \eq{BKMR2}),  choosing the linear and non-linear parts of the BK equation,   in the following form:
\beq\label{DEC:OP}
\mathscr{L}[F]=\dfrac{\pp F}{\pp \Y}+\kappa \dfrac{\pp F}{\pp \zz}-\chi_{0}\left( \dfrac{\pp}{\pp \zz}\right)F+\zz F  + 2\,\Phi\Lb \dY\Rb\,F\,,\quad \mathscr{N_{L}}[F]=\dfrac{\pp }{\pp \zz}\left(\intl_{0}^{\zz} F dz'\right)^{2}
\eeq

The homotopy equation takes the following form:
\beq \label{PRO:HOMO}{\cal H}\Lb p, M_{p}(\zz,b,\dY)\Rb\,\,=\,\,
\mathscr{L}[M_{p}]+p\mathscr{A}_{p}\,\,=\,\,0\eeq

We are looking for  the solution of \eq{DEC:OP}  in the form:
\beq\label{SOL:HOMO}
F\,\equiv\,\,M_{p}(\zz,b, \dY)=\sum_{n=0}^{\infty}M_{n}(\zz,b,\dY)\,p^{n}
\eeq

Plugging \eq{SOL:HOMO} into the homotopy equation (see \eq{PRO:HOMO})
we obtain the following equation for functions $M_n$:

\beq \label{PRO:HOMO1}
\mathscr{L}[M_{n}]+p\mathscr{A}_{n-1}\,\,=\,\,0
\eeq
where 
 $\mathscr{A}_{p}=\sum_{n=0}^{\infty}A_{n}(\zz,b,\dY)p^{n}$    being
\beq\label{NL:HOMO}
A_{n}(\zz,b,\dY)=\sum_{j=0}^{n}\left( \intl_{0}^{\zz}M_{j}(z',b,\dY) \right)\left( \intl_{0}^{\zz}M_{n-j}(z',b,\dY) \right)
\eeq
In the explicit form \eq{PRO:HOMO1} takes the form:
\beq\label{REC}
 \dfrac{\pp M_{n}}{\pp \Y}+\kappa \dfrac{\pp M_{n}}{\pp \zz}=\chi_{0}\left( \dfrac{\pp}{\pp \zz}\right)M_{n}-\zz M_{n}  - 2\,\sum_{j=0}^{n}\Phi_{n - j}\Lb \dY\Rb\,M_j-\dfrac{\pp A_{n-1}}{\pp \zz}
 \eeq
where $\Phi\Lb \dY\Rb \,\,=\,\,\sum_{n} \,\Phi_n\Lb \dY\Rb\,p^n$. As we discuss below,
we start from finding solution at $p=0$ which satisfies \eq{BCINT} at $ z=\xi^A_0$ and \eq{MVF} at $\dY = 0$.  It turns out that  we need to introduce $\Phi_0\Lb \dY\Rb= \h \zeta$. For such $\Phi_0$  \eq{REC} can be rewritten in the form:
\beq\label{REC0}
 \dfrac{\pp M_{n}}{\pp \Y}+\kappa \dfrac{\pp M_{n}}{\pp \zz}=\chi_{0}\left( \dfrac{\pp}{\pp \zz}\right)M_{n}- \Lb \zz\,+\,\zeta\Rb M_{n}  - 2\,\sum_{j=0}^{n-1}\Phi_{n - j}\Lb \dY\Rb\,M_j-\dfrac{\pp A_{n-1}}{\pp \zz}
 \eeq
In the case of $n = 0$ we do not have two last terms of \eq{REC0}.

\section{ 
Solution in the saturation region}

\subsection{ 
Zero iteration (p=0) solution}

For $p=0$  we need to solve the linear equation for $M_0(\zz,b, \dY)$. 
The equation does not depend on  $A(\zz, b, \dY)$ of \eq{AUXFUN} but the term $\zz M_{0}$ in \eq{REC} actually stems from the term $\tilde{N}^2$ in the BK equation.

From \eq{REC0} we see that the  equation for $M_0(\zz,b, \dY)$ takes the form:
\beq\label{M0EQ}
 \dfrac{\pp M_{0}}{\pp\, \dY}+\kappa \dfrac{\pp M_{0}}{\pp \zz}=\chi_{0}\left( \dfrac{\pp}{\pp \zz}\right)M_{0}-(\zz + \zeta) M_{0}
 \eeq

The particular solution  to this equation depends only on $\zz$ and it can be found from the equation:

\beq \label{BKMR4}
\kappa \frac{ d M_0\Lb\zz, b\Rb}{d \zz}\,\,\,=\,\,\,\chi_{0}\left( \dfrac{d}{d \zz}\right)M_0\Lb \zz,  b\Rb \,\,-\,\,\,(\zz\,+\,\zeta) \,M_0\Lb \zz, b\Rb
\eeq

 We are looking for the solution, which has the following form in two kinematic regions, which are shown in \fig{sat}\footnote{ The saturation momentum as well as all scattering amplitudes depends on  impact parameters $b$, which plays a role of external  parameter in our approach. Starting from this place sometimes we will not write $b$ as an argument assuming the implicit dependence, which easily can be recovered. As we have mentioned in the introduction we consider $r \,\ll\,R_N$ and in this case our dependence on $b$  is concentrated in the dependence of the saturation momentum on the impact parameter, which is included in the expression of $z$ (see \eq{z}) and $\zz$ (see \eq{XIT}) in our approach.}

 \bea \label{M0EQ2}
 M_0\Lb \zz, \dY\Rb\,\,=\,\,
 \left\{\begin{array}{l}\,\,M^{I}_0\Lb \zz\Rb \,\,\,\,\,\,\,\,\,\,\mbox{for}\,\,\,\,- \kappa\dY\,\,<\,\,\xt\,\,< \,\xt^A_0;\\ \\
\,\,M^{II}_0\Lb \zz,\dY\Rb  \,\,\,\,\,\mbox{for}\,\,\, \xt\,>\,\xt^A_0;\\  \end{array}
\right.
 \eea
where $M_0^{II}\Lb \zz,\dY\Rb$ is the solution to the general equation (see \eq{M0EQ}).
Note, that we use  the momentum representation instead of the coordinate one, which is shown in \fig{sat}. \eq{XIT} leads to $\xi^A = \ln\Lb r^2 Q^2_A(Y_A, b)\Rb \to \xt^A=
\ln\Lb  Q^2_A(Y_A, b )/k^2_T\Rb$.  The boundary   and initial conditions for $M_0\Lb \zz,\dY\Rb$ in the saturation region  we take the same as for the full equation (see \eq{BKMR2} and \fig{sat}):
 \begin{subequations}    
  \bea 
 \hspace{-0.6cm}  \mbox{region}\,  I: &&
   M^{I}_{0}\Lb\tilde{z}\,=\,\xt_0^A, b\Rb =\,\phi_0(b)   - \h = M_0
        ~~~~~ \phi_0(b)\,\ll\,\h;\,\,\label{M0IC1}\\ 
 \hspace{-0.6cm}  \mbox{region}\,   II: &&  M^{II}_{0}(\zz,\dY=0; b)=-\dfrac{1}{2}\left(1-\exp(-\exp(-\xt))\right),  M_0 = -\dfrac{1}{2}\left(1-\exp(-\exp(-\xt^A_0))\right)~\mbox{for}~\xt >\xt^A_0\,\,\label{M0IC2}
 \eea
  \end{subequations}
Note, that this equation for $\xt\,>\,\xt^A_0$ we obtain directly from McLerran-Venugopalan formula of \eq{MVF},   using \eq{M1}. $\phi_0\Lb b \Rb$ is the value of the scattering amplitude at $\zz =\xt^A_0$. As one can see from these equations, we put the boundary condition of \eq{M0IC1} for the    $M^{I}_0$  while  for $M^{II}_0$ we are using the initial condition of \eq{MVF}. 
  \begin{boldmath}
\subsubsection{Geometric scaling solution  $ M^{I}_0\Lb \zz\Rb$}
\end{boldmath}
We solve \eq{BKMR4}  using the Mellin transform
\beq \label{MT}
M_0 \Lb\tilde{z}, b\Rb= \intl^{\epsilon + i \infty}_{\epsilon - i \infty} \frac{d \gamma}{2 \pi i} e^{\gamma ( \zz + \zeta)} m_0\Lb \gamma, b\Rb
\eeq
where  $m_0\Lb \gamma, b\Rb$ satisfies the equation:
\beq \label{BKMT1}
\Lb \kappa \gamma - \chi\Lb \gamma\Rb + 1/\gamma\Rb\,m_0\Lb \gamma,b\Rb = \frac{d m_0\Lb \gamma, b\Rb}{d \gamma}.
\eeq
The solution for $m_0\Lb \gamma, b\Rb$ takes the following form
\beq \label{BKMT2}
m_0\Lb \gamma\Rb\,\,=\,\,m(0)\exp\Bigg(\intl^\gamma_0  d \gamma' \Big(\kappa \gamma' -  \chi\Lb \gamma'\Rb +1/\gamma' \Big) \Bigg)
\eeq

Taking into account the explicit form of the BFKL kernel given by \eq{GACR} one can re-write \eq{BKMT2} in the form of the product of two Mellin's transforms
\beq \label{BKMT3}
m_0\Lb \gamma\Rb\,\,=\,\,m(0) \exp\Bigg(\kappa \gamma^2/2 -  2 \psi(1)\gamma  \Bigg)\,\Bigg(\frac{\gamma\Gamma( \gamma)  }{ \Gamma(1 - \gamma)}\Bigg) \eeq
 In \eq{BKMT2} and \eq{BKMT3}
$$m(0)=\frac{\phi_{0}(b)-(1/2)}{ C\Lb \zeta\Rb} ~~~~\mbox{where} ~~~C\Lb \zeta\Rb = 
\intl^{\epsilon + i \infty}_{\epsilon - i \infty}\dfrac{d \ga}{ 2 \pi i} e^{( \zeta + 2 \psi(1))\ga}
e^{\h \kappa \ga^2} \frac{\Gamma\Lb 1 +\ga\Rb}{ \Gamma\Lb 1 -\ga\Rb}
. $$

From \eq{BKMT3}  we obtain

\beq\label{SOL01}
	 	M_{0}^{I}(\zz)\, =\, \int_{\epsilon-i\infty}^{\epsilon+i\infty}\dfrac{d\gamma}{2\pi i}e^{\gamma(\dY+\zeta)}\,m(0)\,\left(\dfrac{k_{\perp}}{Q_{s}(Y_{A}, b)}  \right)^{-2\gamma}\dfrac{\Gamma(1+\ga)}{\Gamma(1-\ga)}e^{\kappa\ga^{2}/2}e^{-2\psi(1)\ga}
	 \end{equation}
 
Formula {\bf 6.561.14} of Ref.\cite{RY}    leads to the following expression in the coordinate representation:
 \begin{equation}\label{SOL02}
 	\int_{0}^{\infty}M_{0}^{I}(\zz)J_{1}(k_{\perp}r)dk_{\perp}\, =\, \dfrac{m(0)}{r}\, \int_{\epsilon-i\infty}^{\epsilon+i\infty}\dfrac{
 d\ga}{2\pi i}\left( \dfrac{r^{2}Q_{s}^{2}(Y_{A}, b ) \,e^{\kappa\dY+\zeta-2\psi(1)}}{4} \right)^{\ga}\, e^{\kappa\ga^{2}/2}
 \end{equation}

Taking into account \eq{M1}  and using 
\begin{equation}\label{KERNEL}
	\int_{0}^{\infty}k_{\perp } \,d k_\perp \,J_{1}(k_\perp\, r)J_{1}(k_{\perp} r')\, =\, \dfrac{1}{r}\delta(r-r')
\end{equation}
we reduce \eq{SOL02}  to 
\begin{equation}\label{SOL03}
	N^{I}_{0}(z)\, =\, 1\, +\, 2\, m(0)\, \int_{\epsilon-i\infty}^{\epsilon+i\infty}\dfrac{
		d\ga}{2\pi i}\left( \dfrac{r^{2}Q_{s}^{2}(Y_{A}, b)\,e^{\kappa\dY+\zeta-2\psi(1)}}{4} \right)^{\ga}\, e^{\kappa\ga^{2}/2}
\end{equation}

Introducing  $z=\kappa\dY+\ln(r^{2}Q_{s}^{2}(Y_{A}, b ))$ and $z_{0}=2(\psi(1)+\ln 2)$  we obtain from formula 
 { \bf 7.2.1}  of Ref.\cite{BATEMAN}:
 
  \begin{equation}\label{SOL04}
	N^{I}_{0}(z)\, =\, 1\, +\, \dfrac{4m(0)}{\sqrt{2\pi\kappa}}\, \exp\left(-\dfrac{(z+\zeta-z_{0})^{2}}{2\kappa}  \right)
\end{equation}

 Finally, if we write $\tilde{N}_{0}=\int_{0}^{\infty}rdr\frac{J_{0}(rk_{\perp})}{r^{2}}(1-\Delta_{01})$ with $\Delta_{01}$ presented in \eq{BKLT1}, from \eq{SOL04}   Eq.(4.11) of Ref.\cite{LT} is recovered.

 ~
 
 ~
  \begin{boldmath}
\subsubsection{Transient solution $ M^{II}_0\Lb \zz,\dY\Rb$}
\end{boldmath}
 Assuming $\Phi^{II}_0\Lb \dY\Rb=\zeta/2$, for the region II in \fig{sat} for finding  $M^{II}_0\Lb \zz,\dY\Rb$ we need to solve a general \eq{M0EQ}.   Going to the Mellin transform:
 \beq \label{MIIMT}
 M^{II}_0\Lb \zz,\dY\Rb\,\,=\,\,\intl^{\epsilon + i \infty}_{\epsilon -  i \infty}\frac{d \ga}{2 \pi  i} e^{ ( \zz + \zeta)\, \ga}m^{II}_0\Lb \ga, \dY\Rb
 \eeq
  we can rewrite \eq{M0EQ} in the form:
 \beq\label{GSOL2}
 \dfrac{\pp m^{II}_{0}}{\pp \dY}+( \kappa \ga-\chi(\ga)+1/\gamma )m^{II}_{0}-\dfrac{\pp m^{II}_{0}}{\pp \ga}=  0
 \eeq
 To solve \eq{GSOL2} we introduce $m^{II}_0\Lb \gamma, \dY; b\Rb\,\,=\,\,m_0(\gamma)\,\mu \Lb \gamma, \dY; b\Rb $. For $\mu$ we have:
\beq \label{MU1}
m_0(\gamma)\Bigg(  \dfrac{\pp \mu \Lb \gamma, \dY; b\Rb}{\pp \Y}\,\,-\,\,\dfrac{\pp \mu \Lb \gamma, \dY; b\Rb }{\pp \ga }\Bigg) \,\,=\,\,0\eeq
The solution to \eq{MU1}  is an arbitrary function of  $\dY +\ga$:
\beq \label{MU2}
m^{II}_0\Lb \ga, \dY\Rb \,\,=\,\,m_{0}(\ga) \,\,F\Lb \dY +\ga\Rb
 \eeq 
Function $F$ we can find from the initial conditions of \eq{M0EQ2}  in the region of $\xt\,>\,\xt^A_0$, which has the following form  at $\dY=0$:
 \beq\label{ICMII}
M^{II}_0\Lb \zz,  \dY = 0\Rb=\!\! \int\limits_{\epsilon-i\infty}^{\epsilon+i\infty}\!\!\!\dfrac{d\ga}{2\pi i}e^{\gamma\,(\xt  +  \zeta)}\,m^{II}_{0}(\ga, \dY = 0)\,\,=\,\,\h\Bigg( 1 - \exp\Lb -\,e^{-\xt}\Rb\Bigg) = \intl_{\epsilon-i\infty}^{\epsilon+i\infty}\!\!\!\dfrac{d\ga}{2\pi i}e^{\gamma\,\xt}\,\h  \Gamma(\ga) \eeq
Note, that $\epsilon $ in \eq{ICMII} is in the range: $ -1\,<\,\epsilon\,< \,0$.
 
 Using \eq{BKMT3} for $m_0\Lb \ga\Rb$ we obtain from \eq{ICMII} :
  \beq\label{MII1} 
 F\Lb \ga\Rb \,\,=\,\,-\,\frac{1}{2 \,m(0)} \,\Gamma\Lb - \ga\Rb\,e^{ - \h \kappa\,\ga^2- \zeta\ga  + 2 \psi(1)\,\ga}
 \eeq
 This form of function $F$ leads to
 \beq \label{MII2}
e^{\zeta\,\ga}  m^{II}_0\Lb \ga, \dY\Rb\,\,=\,\,-\,\h \dfrac{\Gamma\Lb 1 + \ga\Rb}{\Gamma\Lb 1 - \ga\Rb}\,\Gamma\Lb- (\ga + \dY)\Rb \,f_1\Lb \dY\Rb e^{ -  \kappa \dY \gamma} \eeq
 with
 \beq \label{MII3}
 f_1\Lb \dY\Rb\,\,=\,\,\exp\Lb - \h \kappa (\dY)^2 - \zeta \dY + 2 \psi(1)\,\dY\Rb
 \eeq
 and with:
 \beq \label{MII4}
M^{II}_0\Lb \zz, b\Rb\,=\,f_1\Lb \dY\Rb\,\int\limits_{\epsilon-i\infty}^{\epsilon+i\infty}\frac{d\ga}{2\pi i}u_{1}^{-\ga} \Lb-\,\h \dfrac{\Gamma\Lb 1 + \ga\Rb}{\Gamma\Lb 1 - \ga\Rb}\,\Gamma\Lb- (\ga + \dY)\Rb\Rb
\eeq
where $u_1= \frac{k_\perp^2}{Q^2_s\Lb \dY, b \Rb}$. Note, that at $\delta\Y>0$ we do not have singularities in choosing $ -1\,< \,\epsilon\,<\,1$.
Using \eq{SOL01} as for 
  $M_{0}^{I}$, we estimate
\begin{equation}\label{MII5}
	\int_{0}^{\infty}M_{0}^{II}(\zz,\dY)\, J_{1}(k_{\perp}\, r)dk_{\perp}\, =\, -\dfrac{1}{2}f_{1}(\dY)\, \int_{\epsilon-i\infty}^{\epsilon+i\infty}\dfrac{d\ga}{2\pi i}\,\left( \dfrac{r^{2}Q_{s}^{2}(Y_{A}, b)}{4} \right)^{\ga}\, \Gamma(-(\ga+\dY))
\end{equation} 
Taking $\ga'=\ga+\dY$ we have
\begin{equation}\label{MII6}
	\int_{0}^{\infty}M_{0}^{II}(\zz,\dY)\, J_{1}(k_{\perp}\, r)dk_{\perp}\, =\, -\dfrac{1}{2}f_{1}(\dY)\, \left( \dfrac{r^{2}Q^{2}_{s}(Y_{A}, b )}{4} \right)^{-\dY}\int_{\epsilon'-i\infty}^{\epsilon'+i\infty}\dfrac{d\ga'}{2\pi i}\,\left( \dfrac{r^{2}Q_{s}^{2}(Y_{A},b )}{4} \right)^{\ga'}\, \Gamma(-\ga')
\end{equation}
It can be checked by direct calculations that $M^{II}_0\Lb \zz, \dY = 0\Rb$ satisfies the initial condition of \eq{ICMII}.
From \eq{MII6}  we obtain
\begin{equation}\label{MII7}
	N_{0}^{II}(r,\dY) =\,1\, -\, f_{1}(\dY)\, \left( \dfrac{r^{2}Q^{2}_{s}(Y_{A}, b)}{4} \right)^{-\dY} \exp\left(-\dfrac{r^{2}Q^2_{s}(Y_{A}, b)}{4}  \right)
\end{equation} 

It worthwhile mentioning that \eq{MII7},  can be trusted only for $z\,>\,0$  while for $z\,<\,0$  we need to use the solution of the linear BFKL equation which in the vicinity of the saturation scale has the form of \eq{VQS}. \eq{M1} allows us to reconstruct $N^{II}_{0}(r, \dY)$ in this kinematic region.

~

~

 \begin{boldmath}
\subsubsection{Matching at $\xi =\xi^A_0$} 
\end{boldmath}

For matching at $\xi =\xi^A_0$
 we rewrite \eq{SOL04} in the form
 \beq \label{MN1}
 N^{I}_0\Lb z\Rb\,\,=\,\,1 \,- \,{\rm D_1} \exp \Lb - \frac{(z \,+\,\zeta\,- z_0)^2}{2\,\kappa}\Rb
 \eeq
 with $z_0 = 2 (\ln2\,+\,\psi(1))$.
 Coefficient $D_1=- 4m(0)/\sqrt{2\,\pi\,\kappa}$ has to be found from the amplitude at $\xi = \xi^A_0$  and can be found from the equation (see \eq{BCINT}):
 \beq \label{MN2}
 {\rm D_1} \exp \Lb - \frac{(\xi^A_0 \,+\,\zeta \,- z_0)^2}{2\,\kappa}\Rb= \exp\Lb -\frac{1}{4} e^{\xi^A_0}\Rb\,\,\equiv\,\,1\,-\,N_0; \eeq
 and the value of $\zeta$ can be found from  the condition (see \eq{BCINT}):
 \beq \label{MN21}
 \frac{d N_0\Lb z\Rb}{d\,z}\Bigg{|}_{z=0} = \bar{\ga} \,N_0
 \eeq
 which translates as the following equation for $\zeta$:
 \beq \label{MN22}
\frac{(\zeta \,-\,z_0)}{\kappa} ( 1 - N_0) = \bar{\ga} N_0
 \eeq

Introducing
\begin{equation}\label{MN5}
	\Phi(z,\xi,b)=(z  +  \zeta - z_0)^{2} - (\xi   +  \zeta -  z_0)^{2}  + \kappa\dfrac{e^{\xi}}{2}
\end{equation}
\eq{MII7}  takes the form:
\begin{equation}\label{MN6}
	N^{II}_{0}(r,b, \dY)=1-\exp\left\lbrace -\dfrac{\Phi(z,\xi,b)}{2\kappa} \right\rbrace
\end{equation}
On the other hands, introducing
\begin{equation}\label{MN7}
	G_{0}(\xi)=\exp\left\lbrace \dfrac{(\xi  +  \zeta - z_0)^{2}}{2\kappa}-\dfrac{e^{\xi}}{4} \right\rbrace
\end{equation}
we can rewrite \eq{MII7}  as follows:
\begin{equation}\label{MN8}
	N^{II}_{0}(\xi, z)=1-G_{0}(\xi)\exp\left\lbrace -\dfrac{(z  +  \zeta -  z_0)^{2}}{2\kappa} \right\rbrace
\end{equation}

\begin{figure}
 	\begin{center}
 	\leavevmode
	\begin{tabular}{c c}
 		\includegraphics[width=9cm]{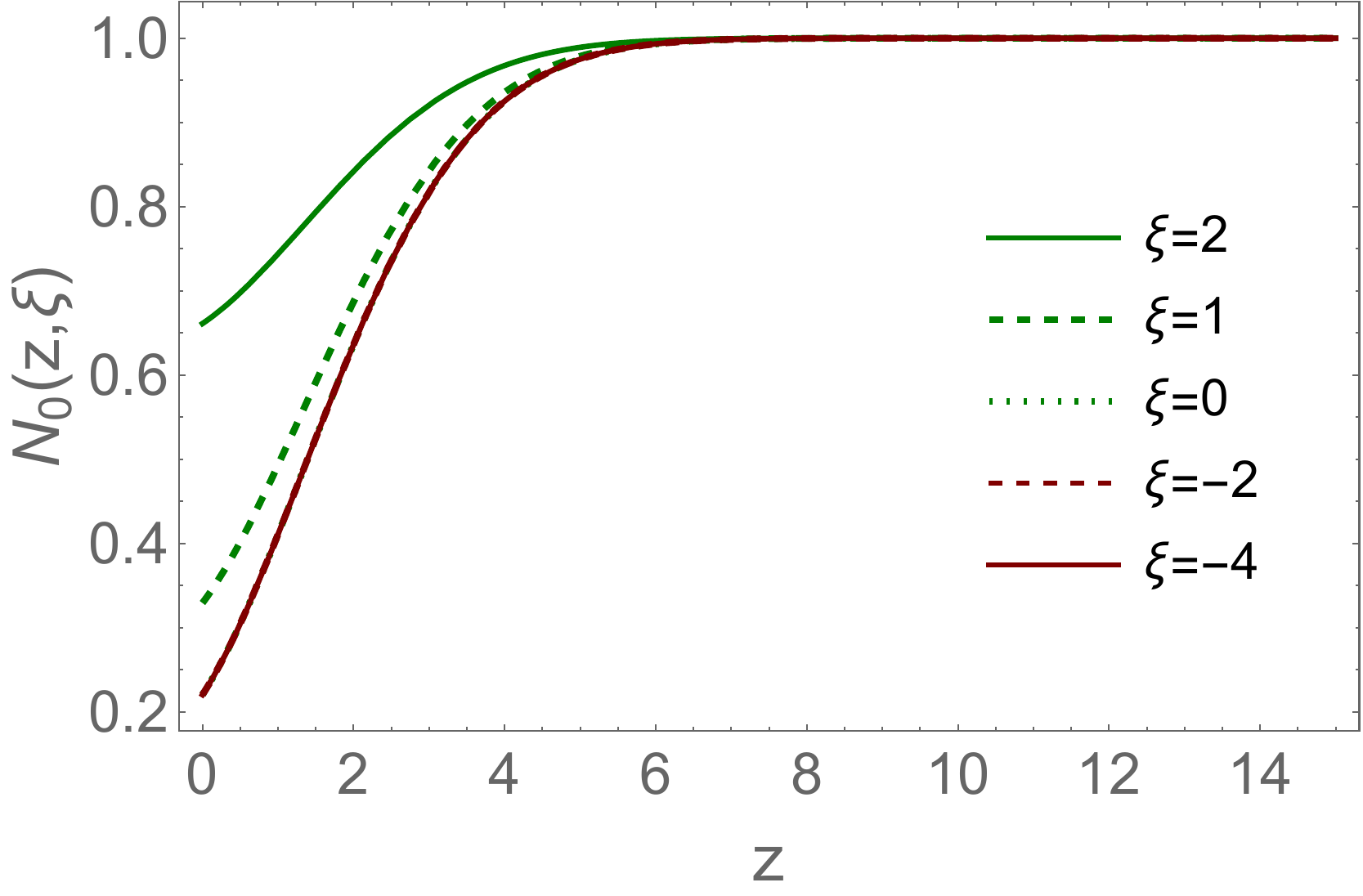}&\includegraphics[width=9cm]{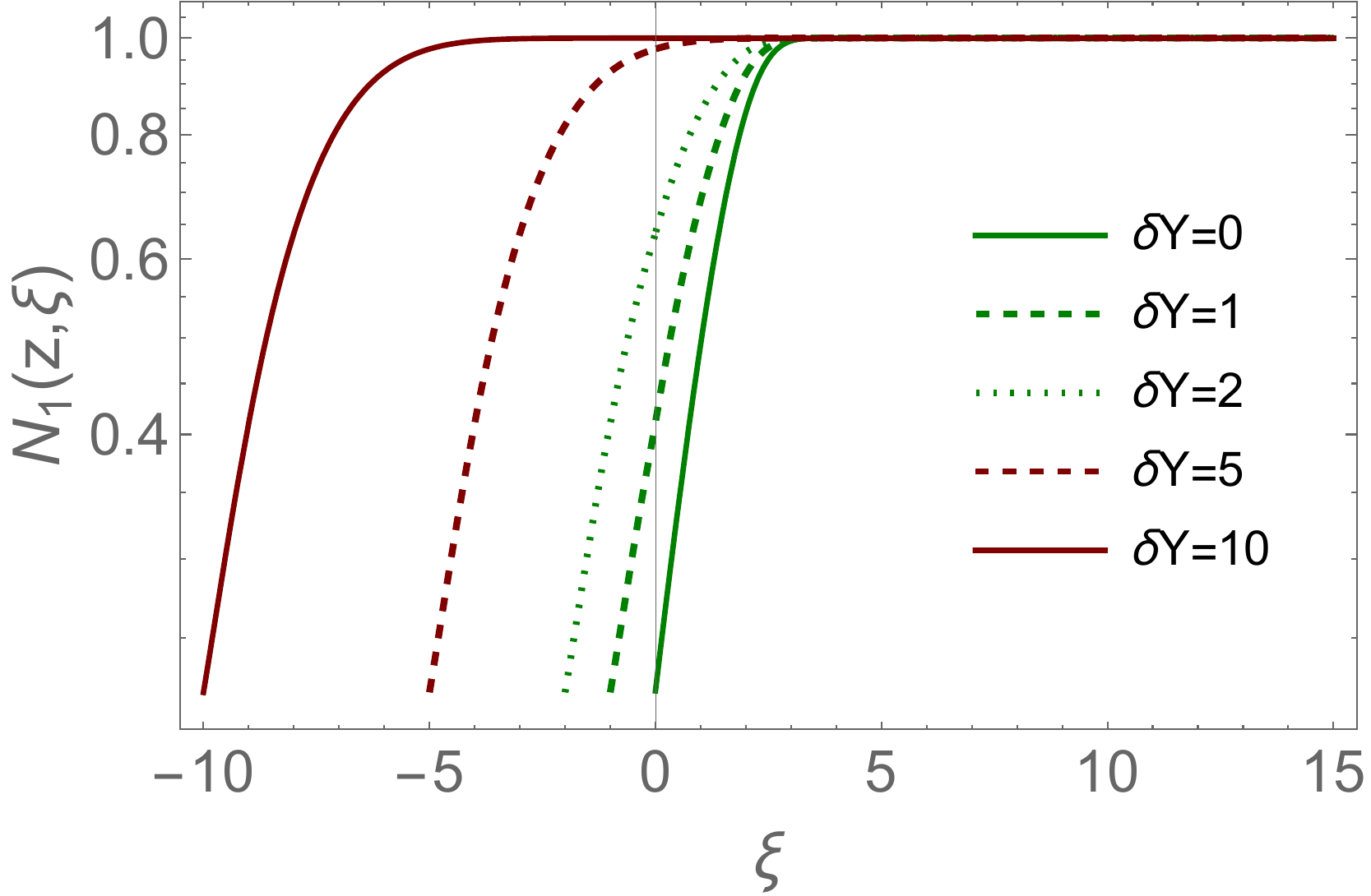}\\
		\fig{n0}-a&\fig{n0}-b\\
		\end{tabular}
		 	\end{center}
 	\caption{ \fig{n0}-a: $N_0(z,\xi) $ versus $z$ at fixed values of $\xt$.
	\fig{n0}-b:   $N_0(z,\xi)$ versus $\xi$ at fixed $\dY$ ($z =  \kappa \dY +\xi$).  
	$N_0 = 0.22;  \\
 \bar{\ga} = 0.63, \, \zeta - z_0 = 0.867, \, C_1= 0.842,  \, \xi^A_0 = 0$.
   }
 	\label{n0}
 \end{figure}

The matching condition for $N^{I}_0$ and $N^{II}_0 $ reads as follows:
\beq \label{MN9}
N^{I}_0\Lb z= \kappa\dY + \xi^A_0\Rb\,\,=\,\,N^{II}_{0}(\xi=\xi^A_0,z= \kappa\dY + \xi^A_0 )
\eeq
From \eq{MN9} we obtain that the equation for $\xi^A_0$ has the form of \eq{XI0}. Hence these two solutions satisfy the matching conditions. In \fig{n0} we present the solution at $p=0$. From the analytical solution as well as from the numerical estimates in \fig{n0} one can see that the geometric scaling behaviour  preserves only in the region I while in the region II we see  large scaling violation.

~

\subsection{First iteration (linear in p)} 

For the first iteration, that takes into account the non-linear corrections we 
 have  the following equation for $M_{1}(\zz,b,\dY)$ (see  \eq{SOL:HOMO}: 
  \beq\label{M1EQ}
\dfrac{\pp M_{1}}{\pp \dY}+\kappa \dfrac{\pp M_{1}}{\pp \zz}=\chi_{0}\left(\dfrac{\pp }{\pp \zz}\right)M_{1}-( \zz +  \zeta)M_{1}\,-\,2\,\Phi_1\Lb \dY\Rb\,M_0-\dfrac{\pp A_{0}}{\pp \zz}
\eeq
 with $A_{0}(\zz,\dY, b)=\Lb \int_{0}^{\zz}M_{0}(z',\dY) dz'\Rb^{2}$ (see \eq{NL:HOMO}).

  $M_{1}(\zz,b,\dY)$ is the solution to linear \eq{M1EQ}, which satisfies the following boundary and initial condition (see \fig{sat}):

   \beq\label{M1IC}
  \mbox{region}\,I: \,\,M_{1}(\zz=\xt^A_0) \,\,=\,\,0;~~~~   
 \mbox{region}\,II:~~~  M_{1}(\zz,b, \dY=0)=\,\,0. \eeq
 
   These boundaries and initial conditions show that both \eq{BCINT} at $z = \xi^A_0$ and \eq{MVF} at $\dY=0$ have been satisfied by $N_0\Lb \zz,\dY\Rb$.

 \begin{boldmath}
\subsubsection{$ M^I_1\Lb \zz \Rb$ and  $ N^I_1\Lb z \Rb$}
\end{boldmath}

In the region I we are looking for the solution which is a function of $\z$.  In addition we assume that $\Phi^{I}_1\Lb\dY\Rb = 0$\footnote{ $\Phi^{I}_1\Lb\dY\Rb = 0$  instead of the arbitrary constant  is taken. Since  such a constant corresponds to redefinition of $d {\cal A}_0/d\zz$, making it is not equal to 0 at $\zz=0$ (see \eq{M1I11}).   $ d {\cal A}_0/d\zz$ takes into account the non-linear corrections from the BK equation and we do not wish to change them.}
Hence, the equation takes the form:
\begin{equation}\label{M1I11}
	\kappa\dfrac{dM_{1}^{I}}{d\zz}\, =\, \chi_{0}\left( \dfrac{d}{d\zz} \right)M_{1}^{I}-(\zz+\zeta)M_{1}^{I}-\dfrac{dA_{0}^{I}}{d\zz},\quad \textrm{with}\quad  \dfrac{dA_{0}^{I}}{d\zz}\, =\,  2M_{0}^{I}\int_{0}^{\zz}M^{I}_{0}(z')dz'
\end{equation}

 Introducing $m_{1}^{I}(\ga,\dY)$ and $a_{0}^{I}(\ga)$ such that
\begin{equation}\label{M1I12}
	M_{1}^{I}(\zz)\, =\, \int_{\epsilon-i\infty}^{\epsilon+i\infty}\dfrac{d\ga}{2\pi i}e^{(\zz+\zeta)\ga}m_{1}^{I}(\ga,\dY),\quad \, \dfrac{dA_{0}^{I}}{d\zz} =\,  \int_{\epsilon-i\infty}^{\epsilon+i\infty}\dfrac{d\ga}{2\pi i}e^{(\zz+\zeta)\ga}\, a_{0}(\ga)
\end{equation}

we reduce \eq{M1I11} to the form:

\begin{equation}\label{M1I13}
	\dfrac{d m_{1}^{I}}{d\gamma}-\left(\kappa\ga-\chi(\ga) +\dfrac{1}{\ga} \right)m_{1}^{I}\, =\,\, a_{0}(\ga)
\end{equation} 

Using variation of parameter, one can see that the general solution of \eq{M1I13}  looks as follows:

\begin{equation}\label{M1I14}
m_{1}^{I}(\ga)\, =\,\left(C_{1}\,+\,\int_{0}^{\ga}d\ga'\dfrac{a_{0}(\ga')}{m_{0}(\ga')}\right)\, m_{0}(\ga)
\end{equation}
In \eq{M1I14} the first term is a general solution of the homogeneous  equation  (see \eq{BKMT1}) and the second one is the particular solution  to the inhomogeneous equation. In \eq{M1I14} 
 $C_{1}$ is chosen such that $M_{1}^{I}(\zz= \xt^A_0)\, =\, 0$.
 
  Since
\begin{equation}\label{M1I15}
	M_{1}^{I}(\zz)\, =\, \dfrac{1}{2}\int_{0}^{\infty}dr'\, J_{1}(r'\, k_{\perp})N_{1}^{I}(r')
\end{equation}
and using the same procedure as for $M_{0}^{I}$ and $M_{0}^{II}$ we obtain
\begin{equation}\label{M1I16}
N_{1}^{I}(z)\, =\,2\,m(0)\, \intl_{\epsilon-i\infty}^{\epsilon+i\infty}\dfrac{d\ga}{2\pi i}e^{(z+\zeta-z_{0})\ga}\, e^{\frac{1}{2}\kappa\ga^{2}}\left( C_{1}\,+\,\int_{0}^{\ga}d\ga'\dfrac{a_{0}(\ga')}{m_{0}(\ga')} \right)
\end{equation}
where
\begin{equation}\label{M1I17}
	C_{1}\, =\,- \left( \intl_{\epsilon-i\infty}^{\epsilon+i\infty}\dfrac{d\ga}{2\pi i}e^{(\xi^A_0\,+\,\zeta-z_{0})\ga}\, e^{\frac{1}{2}\kappa\ga^{2}}\, \int_{0}^{\ga}d\ga'\dfrac{a_{0}(\ga')}{m_{0}(\ga')} \right)/\left(\intl_{\epsilon-i\infty}^{\epsilon+i\infty}\dfrac{d\ga}{2\pi i}e^{(\xi_{0}^{A}+\zeta-z_{0})\ga}\, e^{\frac{1}{2}\kappa\ga^{2}}\right)
\end{equation}

Finally, the solution takes the form
\beq\label{M1I18}
N^I_1\Lb z\Rb\,\,=\,\,C_1\Lb N^I_0\Lb z\Rb\,-\,1\Rb \,+\,\Delta^I_1\Lb z\Rb=\,\,C_1\Lb N^I_0\Lb z\Rb\,\,-\,\,1\Rb \,+\,\intl_{\epsilon-i\infty}^{\epsilon+i\infty}\dfrac{d\ga}{2\pi i}e^{(z+\zeta-z_{0})\ga}\,m(0) e^{\frac{1}{2}\kappa\ga^{2}}\int_{0}^{\ga}d\ga'\dfrac{a_{0}(\ga')}{m_{0}(\ga')}
\eeq
whereas an estimative about their qualitative behavior (type of contribution) can be expressed as follows
\begin{equation}\label{ESTI:SOL1}
	N_{1}^{I}(z)\, \sim\, -\frac{c_{1}^{I}(\zeta)\, m(0)}{2\kappa\sqrt{2\pi\kappa}}\, (z-z_{0}^{A})\, \Delta_{01}(z)
\end{equation}
with $c_{1}^{I}(\zeta)\, =\, -2\intl_{\epsilon-i \infty}^{\epsilon+i \infty}\frac{d\ga}{2\pi i}\, \frac{d\ga}{2\pi i}e^{\zeta\ga}\frac{1}{\ga}m_{0}^{I}(\ga)$, being fulfilled $\vert N_{1}^{I}(z)\vert\ll N_{0}^{I}(z)$. See Appendix \ref{APPENDIX:A} for more details.

In our approach we have $\Delta_{01}(z)\, =\, e^{-\frac{1}{2\kappa}(z+\zeta-z_{0})^{2}}$, with $\dfrac{1}{\kappa}(\zeta-z_{0})\, =\, \frac{N_{0}}{(1-N_{0})}\, \overline{\ga}$, see \eq{MN22}. Therefore, \eq{ESTI:SOL1}, can be rewritten as
\begin{equation}\label{ESTI:SOL1:B}
	N_{1}^{I}(z)\, \sim\, \exp\left( -\dfrac{z^{2}}{2\kappa} \right)\times \exp\left(-\left(\frac{N_{0}}{1-N_{0}}\right)\, \overline{\ga}\, z\, +\, \ln(z-z_{0}^{A})  \right)
\end{equation}
with $\overline{\ga}\, =\, 1-\ga_{cr}$.
~

~

 \begin{boldmath}
\subsubsection{$M_{1}^{II}(\zz,\dY)$ and $N_{1}^{II}(z,\dY)$}
\end{boldmath}

  In the region II the solution depends on two variables: $\zz$ and $\dY$. The term $ \dfrac{\pp A^{II}_{0}}{\pp \zz}$ in the master equation  has the form:
\begin{equation}\label{M1I21}
\dfrac{\pp A^{II}_{0}}{\pp \zz}\, =\, 2M^{II}_{0}(\zz,\dY)\left(\la_{0}(\dY)\, -\,  \int_{\zz}^{\infty}d\zz'M_{0}^{II}(\zz',\dY) \right)
\end{equation}
where 
\begin{equation}\label{M1I22}
\la_{0}(\dY)\, =\, \int_{0}^{\kappa\dY}d\zz'M_{0}^{I}(\zz')\, +\, \int_{\kappa\dY}^{\infty}d\zz'M^{II}_{0}(\zz',\dY)
\end{equation}
Introducing the Mellin transforms  for $m_{1}^{II}(\ga,\dY)$ and $a_{0}^{II}(\ga,\dY)$ such that
\begin{equation}\label{M1I23}
	M_{1}^{II}(\zz,\dY)\, =\, \int_{\epsilon-i\infty}^{\epsilon+i\infty}\dfrac{d\ga}{2\pi i}e^{(\zz+\zeta)\ga}m_{1}^{II}(\ga,\dY),\quad  2\, M_{0}^{II}(\zz,\dY)\int_{\zz}^{\infty}d\zz'M^{II}_{0}(\zz',\dY)\, =\,   \int_{\epsilon-i\infty}^{\epsilon+i\infty}\dfrac{d\ga}{2\pi i}e^{(\zz+\zeta)\ga} a_{0}^{II}(\gamma,\dY)
\end{equation}	
Choosing $\Phi_{1}^{II}(\dY)=-\la_{0}(\dY) \,+\,\tilde{\Phi}_{1}^{(II)}\Lb \dY\Rb$ and  repeating the same procedure as in \eq{M1I14},
we reduce  \eq{M1EQ} to the form:
 \beq\label{M1I24}
 \dfrac{\pp m_{1}}{\pp \dY}+( \kappa \ga-\chi(\ga)+1/\gamma )m_{1}-\dfrac{\pp m_{1}}{\pp \ga}\,+\,2\, \tilde{\Phi}^{(II)}_1\Lb \dY\Rb m_0=  \,\,\,\,a^{II}_0(\ga,\dY)
 \eeq
 In \eq{M1I23} we consider both $M_{1}\Lb\zz,\dY\Rb$ and  $A_{0}\Lb\zz,\dY\Rb$   being equal to zero for negative $z$.

To solve \eq{M1I11} we introduce $m_1\Lb \gamma, \dY\Rb\,=\,-\,2\, \intl^{\dY}_0\!\!\!d \dY'\tilde{\Phi}^{(II)}_1\Lb \dY'\Rb m_0(\ga, \dY)  +  \tilde{m}_1\Lb \ga, \dY\Rb $. Plugging this form of $m_1\Lb \gamma, \Y\Rb $ in \eq{M1I15} we obtain
 \bea \label{M1EQ10} 
 &&-\,2\, \intl^{\dY}_0\!\!\!d \dY'\tilde{\Phi}^{(II)}_1\Lb \dY'\Rb\underbrace{ \Bigg\{ \frac{\pp m^{II}_0\Lb \ga, \dY\Rb}{\pp \dY} +( \kappa \ga-\chi(\ga)+1/\gamma )m^{II}_{0}\Lb \ga, \dY\Rb-\dfrac{\pp  m^{II}_{0}\Lb \ga, \dY\Rb}{\pp \ga} \Bigg\}}_{ \mbox{ =  0 ( see \eq{GSOL2})}} \nn\\ 
 &&+ \dfrac{\pp \tilde{m}_{1}}{\pp \dY}+( \kappa \ga-\chi(\ga)+1/\gamma )\tilde{m}_{1}-\dfrac{\pp \tilde{m}_{1}}{\pp \ga}  =  \,\,\,\,a^{II}_0(\ga,\dY) 
 \eea
 Searching a solution in the following form: $\tilde{m}_1\Lb \ga,\dY\Rb\,\,=\,\,m_0(\gamma)\,\mu_1 \Lb \gamma, \dY; b\Rb $,   we have for $\mu_1$:
\beq \label{MU01}
m_0(\gamma)\Bigg(  \dfrac{\pp \mu_1 \Lb \gamma, \dY; b\Rb}{\pp \Y}\,\,-\,\,\dfrac{\pp \mu_1 \Lb \gamma,\dY; b\Rb }{\pp \ga }\Bigg) \,-\, a^{II}_{0}(\ga, \dY)\,\,=\,\,0; ~~~ \dfrac{\pp \mu_1 \Lb \gamma, \Y; b\Rb}{\pp \Y}\,\,-\,\,\dfrac{\pp \mu_1 \Lb \gamma, \dY; b\Rb }{\pp \ga } = \,\,\,\dfrac{ a^{II}_{0}(\ga, \dY)}{m_0(\ga)}\eeq

The solution to \eq{MU01} can be found in the $\zz$ - representation, introducing
\beq \label{MU01B}
\tilde{\mu}_1\Lb \zz,\dY\Rb\,\,=\,\,\intl^{\epsilon + i \infty} _{\epsilon - i \infty} \frac{d \ga}{2 \pi i} e^{ \ga\,\zz} \frac{ \tilde{m}_1\Lb \ga, \dY\Rb}{m_0\Lb \ga\Rb};~~~~~~~\tilde{A}\Lb \zz,\dY\Rb
\,\,=\,\,\,\intl^{\epsilon + i \infty} _{\epsilon - i \infty} \frac{d \ga}{2 \pi i} e^{ \ga\,\zz} \frac{a^{II}_0\Lb \ga, \dY\Rb}{m_0\Lb \ga\Rb};
\eeq
\eq{MU01} takes the form
\beq \label{MU02}
\frac{\pp \tilde{\mu}_1\Lb \zz,\dY\Rb}{\pp \,\dY} \,\,+\,\,\zz\,\tilde{\mu}_1\Lb \zz,\dY\Rb=\tilde{A}\Lb \zz,\dY\Rb
\eeq

The general solution to \eq{MU02} can be found in  the form: $\tilde{\mu}_1\Lb \zz,\dY\Rb\,\,=\,\,e^{-\,\zz\,\dY} H\Lb \zz,\dY\Rb$, with the following equation for $H$:
\beq \label{MU03}
\frac{\pp H\Lb \zz,\dY\Rb}{\pp \,\dY} =  e^{\zz\,\dY}\tilde{A}\Lb \zz,\dY\Rb
\eeq
Finally, the solution, which satisfies the initial condition at $\dY = 0$ (see \eq{M1IC}), has the form:
\beq \label{MU04}
\tilde{\mu}_1 \Lb \zz,\dY\Rb  =  e^{-\zz\,\dY} \,\intl_0^{\dY} d t e^{ \zz t} \tilde{A}\Lb \zz,\,t\Rb
\eeq
 For the  Mellin image of $\tilde{\mu}_1 \Lb \zz,\dY\Rb$ we have:
 \beq \label{MU05}
\mu_1\Lb \ga,\dY\Rb \,\,=\,\,  \intl^{\dY}_0 d t \frac{a^{II}_0\Lb \ga + \dY - t,  t\Rb}{m_0\Lb \ga + \dY - t\Rb} 
\eeq 
 
 Hence,  the general solution to \eq{M1I24}, which satisfies the  initial  and boundary conditions of \eq{M1IC} takes the following form:  
\beq \label{MU2B}
m^{(II)}_1\Lb \ga, \dY\Rb\,\,=\,\,\,-\,2\, \intl^{\dY}_0\!\!d t\tilde{\Phi}^{II}_1\Lb t \Rb m_0(\ga, \dY)\,\,+\,\, m_0(\ga)\intl^{\dY}_0 d t   \frac{ a^{II}_0\Lb \ga + \dY - t, t\Rb}{m_0\Lb \ga+ \dY - t\Rb}\eeq

   The general solution for $N_{1}^{II}\Lb r, \dY\Rb$ has the following form
  
\bea \label{NII1}
&&N^{II}_1\Lb z,\dY\Rb\,\,\,=\,\,
\,\,2\,m(0)\intl^{\epsilon +i \infty}_{\epsilon - i \infty} \!\!\!\!
\dfrac{d \,\ga}{2\,\pi\,i}\,e^{(z +\zeta - z_0) \,\ga} e^{\h \kappa \ga^2}\,\intl^{\dY}_0 d t   \frac{ a^{II}_0\Lb \ga + \dY - t, t\Rb}{m_0\Lb \ga+ \dY - t\Rb} + I\Lb z, \dY\Rb
   \eea
where $ I\Lb z, \dY\Rb \,=\,-\,\intl^{\dY}_0 d\dY' \tilde{\Phi}^{II}_1\Lb \dY'\Rb\Big( N^{II}_0\Lb z,\dY\Rb \,-\,1\Big)$. The developments for the analysis of the behavior of the contributions in \eq{NII1} are presented in the appendix section, see \eq{EXP:SOL:P1}. Our analysis led us to the following approximation
\begin{equation}\label{NII10}
	N_{1}^{II}(z,\dY)\, =\, h(\dy)\, \exp\left( -\frac{1}{2}(z+\zeta-z_{0})\dY \right)\, \exp\left( -\frac{1}{2} (\xi+\zeta-z_{0})\dY\right)\, \exp\left( -\dfrac{r^{2}Q^{2}_{s}(Y_{A}, b )}{4}\right)\, (1+o(1))
\end{equation}
with $h(\dY)\, =\, -\int_{0}^{\dY}d\dY'\tilde{\Phi}_{1}^{II}(\dY')$.
~

~

 \begin{boldmath}
\subsubsection{Matching at $\xi =\xi^A_0$} 
\end{boldmath}

The matching condition at $\xi = \xi^A_0$ is
\beq \label{MM11}
N^{I}_1\Lb z\Rb\,\,=\,\,N^{II}_1\Lb z,\dY\Rb~~~\mbox{for}~~~z=  \kappa\,\dY + \xi^A_0
\eeq
Plugging in \eq{MM11} the solutions presented in  \eq{M1I16} and \eq{NII1} one can see 
that \eq{MM11} takes the following form
\beq \label{MM12}
N^{I}_1\Lb z = \xi^A_0 +\kappa \dY\Rb\,\,= \,\, -2 \, m(0)\,\intl^{\epsilon + i\infty}_{\epsilon - i \infty} \!\!
 \dfrac{d \,\ga}{2\,\pi\,i}e^{(\xi^A_0 +\kappa \dY)\ga} e^{\h \kappa\,\ga^2}\,\,\intl^{\dY}_0 d t   \frac{ a^{II}_0\Lb \ga + \dY - t, t\Rb}{m_0\Lb \ga+ \dY - t\Rb}\,+\,I\Lb \xi^A_0 +\kappa \dY , \dY\Rb\eeq

We need to recall, that $ a_{0}(\ga', \dY)$ has different forms for $M^{I}_1\Lb \zz\Rb$ and for $M^{II}_1\Lb \zz, \dY\Rb$. For the region I  $a^{I}_{0}(\ga)$ it is a Mellin image of $ 2\, M^{I}_0\Lb \zz\Rb \intl^{\zz}_0 d \zz'\,M^{I}_0\Lb \zz'\Rb$, while the region II $a^{II}_{0}(\ga, \dY)$  it is a Mellin image of \eq{M1I21} and \eq{M1I22}. Actually the contribution of $M^{I}_0$
 leads to the change of function $\Phi^{II}_1\Lb \dY\Rb$, and, therefore

\beq \label{MM120}
a_0^{II}\Lb \ga,\dY\Rb\,\,=\,\,\intl^{\infty}_0 d\,\zz \,e^{ -\ga\,\zz} \, 2\,M^{II}_0\Lb \zz,\dY\Rb \,\,\intl^{z}_{\kappa\,\dY}d \zz'\, M^{II}_0\Lb \zz',\dY\Rb
\eeq

Function $\Phi\Lb \dY\Rb$ has to be found from the following equation:
\beq
 \label{MN14}
\intl^{\dY}_0 d\dY' \Phi^{II}_1\Lb \dY'\Rb \Big( N^{II}_0\Lb \xi^A_0 + \kappa \dY, \dY\Rb \,-\,1\Big)
=  
N^{I}_1\Lb \xi^A_0 + \kappa\,\dY\Rb \,+\,2\,\intl^{\epsilon + i \infty}_{\epsilon - i \infty} 
\dfrac{d \,\ga}{2\,\pi\,i}e^{(\xi^A_0 +\kappa \dY)\ga}\,e^{\h \kappa\,\ga^2} \intl^{\dY}_0 d t   \frac{ a^{II}_0\Lb \ga + \dY - t, t\Rb}{m_0\Lb \ga+ \dY - t\Rb}
\eeq
 \begin{boldmath}
\subsubsection{Numerical estimates} 
\end{boldmath}


{\boldmath $N^{I}_1\Lb z \Rb$:}

For numerical estimates we use \eq{M1I16} calculating the integral over $\ga$ along the imaginary axis.  The integral is concentrated in vicinity of small $\ga$  due to the factor $\exp\Lb - \h \kappa \Lb {\rm Im}\, \ga\Rb^2\Rb$.  The second term in \eq{M1I16} can be rewritten in more convenient form for numerical estimates, viz.:

\beq \label{NE1}
-2\!\!\intl^{\epsilon + i \infty}_{\epsilon - i \infty} \!\!\!\frac{d \ga}{2\,\pi\,i}e^{( z + \zeta  -z_{0})\,\ga} e^{\h \kappa \ga^2}\,\intl^{\ga}_{0} d \ga' \dfrac{ a_{0}(\ga')}{m_0(\ga')}\,\,=\,\,-2 \intl^{ \infty}_{ 0}\!\frac{d \nu}{\pi} {\rm Re}\Bigg( e^{ i( z + \zeta  -z_{0})\,\nu}\,i\,\int^\nu_0\!\!\! d t \exp\Lb - \h \kappa\,\nu ( 2 \nu - t)\Rb a_0\Lb \nu - t\Rb\Bigg)
\eeq

One can see from \eq{NE1} that integral over $\nu$ is well converged  at $\nu $ close to 0.

\begin{figure}
 	\begin{center}
 	\leavevmode
	\begin{tabular}{cc}
	\includegraphics[width=9cm]{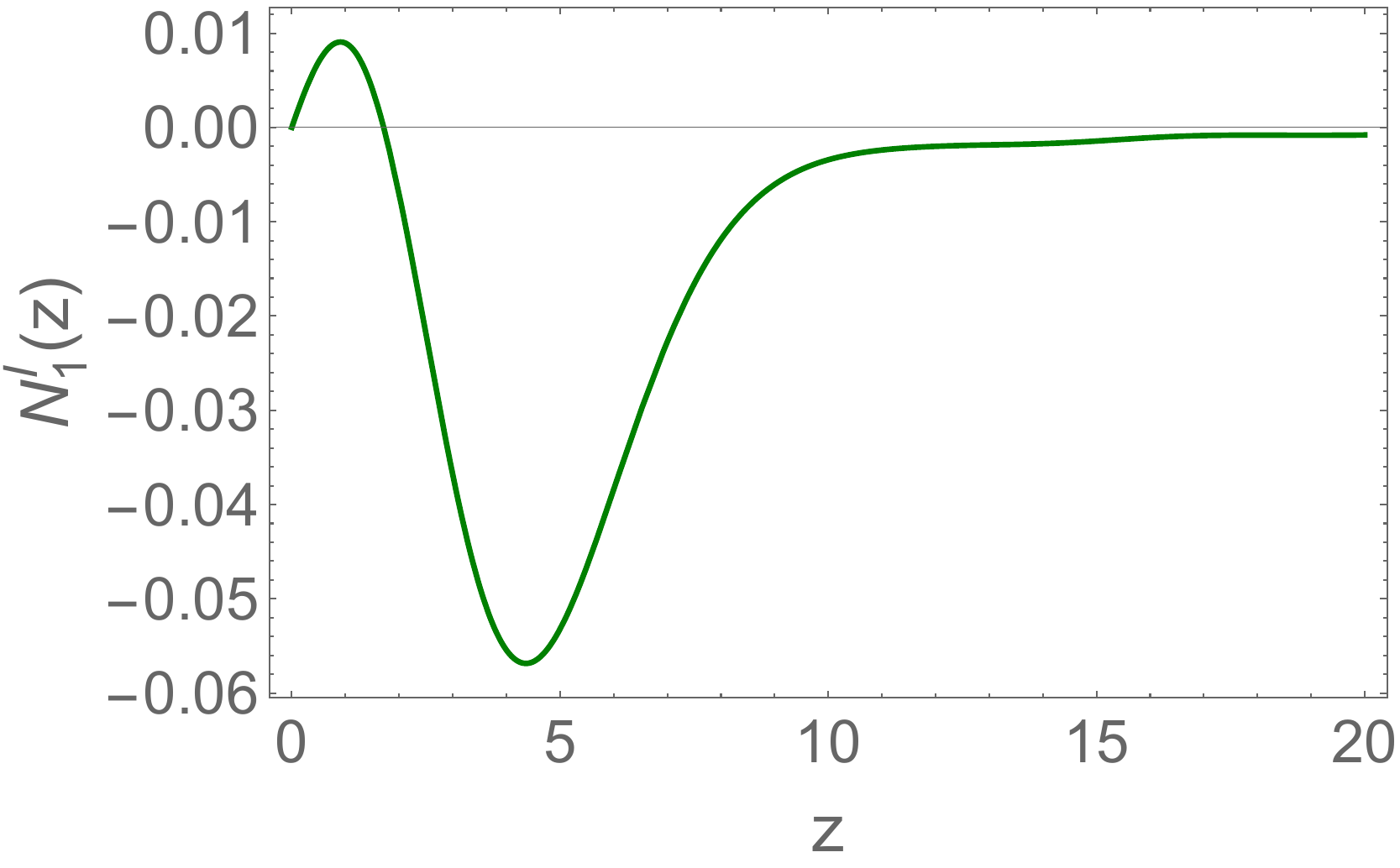}&\includegraphics[width=9cm]{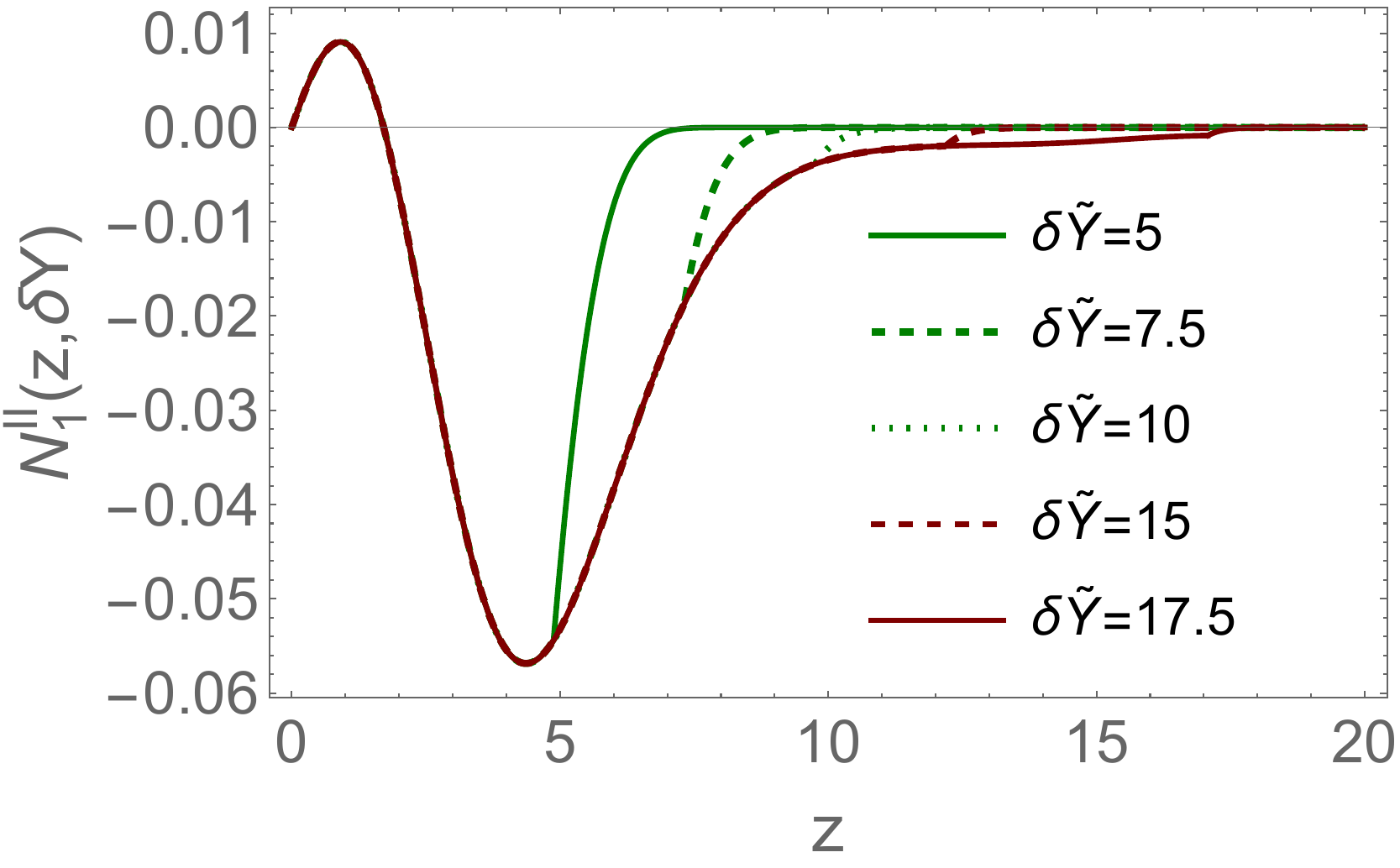} \\
	\fig{n1}-a & \fig{n1}-b\\
	\end{tabular}
		\end{center}
 	\caption{ \fig{n1}-a:  $N^{I}_1\Lb z \Rb$ versus $z$. \fig{n1}-b: $N_1\Lb z, \delta Y\Rb$ versus $z$ at fixed $\dY = \bas (Y - Y_A)$.
The value of $\xi^A_0= 0$. }
 	\label{n1}
\end{figure}

In \fig{n1}-a  we plot the numerical estimates of $N^{I}_1\Lb z\Rb$  for $\xi^A_0=0$. One can see that the value of $N^{I}_1$ turns out to be very small. Hence, we see that out homotopy approach leads to a regular way to take into account the non linear corrections. We can recall that our approach consists of two steps. First we solve the linearized version of the non linear evolution equation at large values of $z$ (see Ref.\cite{LT}). Second,  we calculate the corrections to this solution, using the homotopy approach. The small values of $N^{I}_1$ shows that homotopy approach gives the efficient way to take into account the shadowing corrections.

~

{\boldmath $N^{II}_1\Lb z,\dY \Rb$:}

In \fig{n1}-b we plot the first correction to the scattering amplitude. It should be noted that we did not face any difficulties in estimating the integrals over $\ga$ for $N^{II}_1\Lb z,\dy\Rb$, since the factor $\exp\Lb + \h \kappa \,\ga^2\Rb$ provides the good convergence of the integrals at our imaginary values of $\ga$. This figure shows that at $ z \leq \dY$ we have the geometric scaling behaviour of the scattering amplitude while for $z > \dY$ the amplitude started to depend on $\dY$ violating the geometric scaling behaviour.

 
~

\begin{boldmath}
\subsection{General case $p=n$} 
\end{boldmath}
The general equation for $M_n$ has been written in \eq{REC} with $A_{n-1}$ defined in \eq{NL:HOMO}. The boundary and initial condition for all $M_n$ are the same as for $M_1$ and they are given by \eq{M1IC}.  
 In the same way as has been  done in the first iteration, the analysis of the corrections should be developed in two different regions.

~

~
\begin{boldmath}
	\subsubsection{$M^{I}_n\Lb \zz \Rb$ and $N^{I}_n\Lb z\Rb$} 
\end{boldmath}

Assuming $\Phi^{I}_{j}(\dY)=0$ for $j\geq 1$, the contribution is obtained solving:
\begin{equation}\label{GEO:SCAL:GEN}
	\kappa\, \dfrac{dM_{n}^{I}}{d\zz}\, =\, \chi_{0}\left(\dfrac{d}{d\zz}  \right)M_{n}^{I}-(\zz-\zeta)\, M_{n}^{I}-\dfrac{dA_{n-1}^{I}}{d\zz},\quad \textrm{with}\, \dfrac{dA_{n-1}^{I}}{d\zz}\, =\, 2\, \sum_{j=0}^{n-1}M^{I}_{j}(\zz)\, \int_{0}^{\zz}M^{I}_{n-1-j}(z')dz'
\end{equation}
Introducing  Mellin transforms again
\begin{equation}\label{TRAN:MELLIN:GEN}
	M_{n}^{I}(\zz)\, =\, \int_{\epsilon-i\infty}^{\epsilon+i\infty}\dfrac{d\ga}{2\pi i}\, e^{(\zz+\zeta)\ga}\, m_{n}^{I}(\ga),\quad \dfrac{dA_{n-1}}{d\zz}\, =\, \int_{\epsilon-i\infty}^{\epsilon+i\infty}\dfrac{d\ga}{2\pi i}\, e^{(\zz+\zeta)\ga}\, a_{n-1}^{I}(\ga)
\end{equation}
we obtain the ordinary differential equation  for $m_{n}^{I}$ with the same structure  as \eq{M1I13}, with non homogeneous  term $-a_{n-1}^{I}(\ga)$, which depends  recursively on  the previous solutions.

Using a similar procedure as for $N_{1}^{I}(z)$ (see Appendix A),  we obtain
\begin{equation}\label{ODE:GEN:R1}
	\kappa\dfrac{dN_{n}^{I}}{d\zz}\, +\, (z+\zeta-z_{0})\, N_{n}^{I}\, = \,-\, \underbrace{ \int_{\epsilon-i\infty}^{\epsilon+i\infty}\dfrac{d\ga}{2\pi i}\, u^{-\ga}e^{2\psi(1)\ga}\, a_{n-1}^{I}(\ga)\dfrac{\Gamma(1-\ga)}{\Gamma(1+\ga)}}_{H^{I}_{n}(z)}
\end{equation}
where  $u=e^{-(z+\zeta-z_{0})}$. \eq{ODE:GEN:R1} 
satisfies    the condition $N_{n}^{I}(z_{0}^{A})\, =\, 0$, with an approximate contribution given by
\begin{equation}\label{ESTI:SOLN}
	N_{n}^{I}(z)\, \sim\, -\frac{c_{n}^{I}(\zeta)\, m(0)}{2\kappa\sqrt{2\pi\kappa}}\, (z-z_{0}^{A})\, \Delta_{01}(z)
\end{equation}
for some constant $c_{n}(\zeta)$ which depends on the previous solutions: $m_{j}^{I}(\ga)$. As we discuss in the appendix, this contribution is small, confirming our numerical estimates.

~

~

\begin{boldmath}
	\subsubsection{$M^{II}_n\Lb \zz, \dY \Rb$ and $N^{II}_n\Lb z, \dY\Rb$} 
\end{boldmath}
The nonhomogeneous term takes the form:
\begin{equation}\label{NOHOMOgen}
\dfrac{\pp A^{II}_{n-1}}{\pp \zz}\, =\, 2\sum_{j=0}^{n-1}\, M^{II}_{j}(\zz,\dY)\, \int_{0}^{\zz}d\zz'\, M^{II}_{n-1-j}(\zz',\dY)
\end{equation}
The differences in the expressions with respect to the regions are treated considering
\begin{equation*}
	\int_{0}^{\zz}d\zz'\,M_{l}(\zz',\dY) \, =\, \la_{l}(\dY)\, +\, \int_{\zz}^{\infty}d\zz'M^{II}_{l}(\zz',\dY),
\end{equation*} 
with $ \la_{l}(\dY)\, =\,  \int_{0}^{\infty} d\zz'\, M_{l}(\zz',\dY)$. Writing $\Phi_{l+1}^{II}\, =\, -\, \la_{l}(\dY)\, +\, \tilde{\Phi}_{l+1}^{II}(\dY)$, we have
\begin{equation*}
	-2\sum_{j=0}^{n-1}\Phi_{n-j}^{II}(\dY)\, M_{j}^{II}(\zz,\dY)\, -\, \dfrac{\pp A_{n-1}}{\pp \zz}\, =\, -2\, \sum_{j=0}^{n-1}\tilde{\Phi}_{n-j}^{II}(\dY)\, M^{II}_{j}(\zz,\dY)\, +\, \underbrace{2\sum_{j=0}^{n-1}M_{j}^{II}(\zz,\dY)\, \int_{0}^{\zz}d\zz'M^{II}_{n-1-j}(\zz',\dY)}_{\tilde{A}_{n-1}(\zz,\dY)}
\end{equation*}
\eq{REC} for the Mellin images takes the form:
\beq\label{MN1:GEN}
\dfrac{\pp m_{n}\Lb \ga,\dY\Rb}{\pp \dY}+( \kappa \ga-\chi(\ga)+1/\gamma )m_{n}\Lb \ga,\dY\Rb-\dfrac{\pp m_{n}\Lb \ga,\dY\Rb}{\pp \ga}\,+\,2\,\sum_{j=0}^{n-1}\tilde{\Phi}^{II}_{n - j}\Lb \dY\Rb\,m_j\Lb \ga,\dY\Rb=  \,\,\,\,a_{n - 1}^{II}(\ga,\dY)
\eeq
where $\tilde{A}_{n-1}=\int_{\epsilon-i\infty}^{\epsilon+i\infty}\frac{d\ga}{2\pi i}e^{(\zz+\zeta)\ga}a_{n-1}^{II}(\ga,\dY)$. 

Note, that for $n\geq 2$ we need to consider extra  contributions in comparison with \eq{M1EQ10}, 
since the solution for each $p=1,\dots, n-1$ satisfies a non-homogeneous equation.   Thus, from the linearity of the equation, we can solve \eq{MN1:GEN} considering a superposition of non-homegeneous contributions, i.e, $b_{n}(\ga,\dY)=-2\sum_{j=0}^{n-1}\tilde{\Phi}_{n-j}(\dY)m_{j}(\ga,\dY)\, +\, a_{n-1}^{II}(\ga,\dY)$. 
Writing $m_{n}(\ga,\dY)\, =\, m_{0}(\ga)\, \mu_{n}(\ga,\dY)$ we obtain the following equation for $\mu_n$:
\beq \label{MN2.GEN}
\dfrac{\pp \mu_n \Lb \gamma, \dY\Rb}{\pp \dY}\,\,-\,\,\dfrac{\pp \mu_n \Lb \gamma, \dY\Rb }{\pp \ga } = \,\,\,\dfrac{ b_{n}(\ga, \dY)}{m_0(\ga)}\eeq

Following similar procedure as in  \eq{NII1}, we see that  the solution in coordinate space has  the form:
\bea \label{SOLREII:GEN}
&&N^{II}_n\Lb z,\dY\Rb\,\,\,=\,\,
\,\,2\,m(0)\intl^{\epsilon +i \infty}_{\epsilon - i \infty} \!\!\!\!
\dfrac{d \,\ga}{2\,\pi\,i}\,e^{(z +\zeta - z_0) \,\ga} e^{\h \kappa \ga^2}\,\intl^{\dY}_0 d t   \frac{ b_{n}\Lb \ga + \dY - t, t\Rb}{m_0\Lb \ga+ \dY - t\Rb}
\eea
Finally, plugging  $b_{n}(\ga,\dY)=-2\sum_{j=0}^{n-1}\tilde{\Phi}_{n-j}(\dY)m_{j}(\ga,\dY)\, +\, a_{n-1}^{II}(\ga,\dY)$ into \eq{SOLREII:GEN} we obtain  the integral representation of the solution from the recursive definition for each $m_{j}(\ga,\dY)$.

~

~

\section{Conclusions}

In this paper we developed the homotopy approach for solving the non-linear evolution Balitsky-Kovchegov equation\cite{BK}. The solution consists of two steps. First, we solved the linearized version of the BK equation \cite{LT} in the momentum space   deep in the saturation region.  We found that this solution has the geometric scaling behaviour for $\xi \leq \xi^A_0$ (see \fig{sat}). For $\xi \,>\,\xi^A_0$, we observe the strong violation of the geometric scaling behaviour in the saturation region. This solution satisfies the boundary and initial conditions which are given  perturbative QCD approach  for $r\,Q_s < 1$ (see \eq{BCINT})  and  by McLerran-Venugopalan formula of \eq{MVF} for $Y = Y_A$. 

In the second step  of our approach we are  taking into account the remaining part of the non-linear correction  that have not been  included in the linearized form of  BK equation. It turns out that these corrections are rather small indicating that our procedure gives a self consistent way to account them.

 We believe that we suggest the method of finding solution, which allow us to treat the most essential part of the scattering amplitude analytically. The numerical part of the calculations is expressed through well converged integrals and can be easily estimated.  However, in this paper we have not investigated the corrections $N^{I}\Lb z\Rb$ and $
 N^{II}\Lb z,\dY\Rb$ in full using that dependence of $z$ of ${A}_0(z,\dY)$ is determined by the contribution of $M^{I}_0$ and $M^{II}_0$.  These study would make the paper long and unreadable. Hence, we decided to put this in a separate paper with more mathematical content, which will be prepared shortly.
 
\section{Acknowledgements}
   We thank our colleagues at Tel Aviv University and UTFSM for
 encouraging discussions. CC thanks the department of particle physics of  Tel Aviv University for hospitality and support during his stay. This research was supported  by 
 ANID PIA/APOYO AFB180002 (Chile) and  Fondecyt (Chile) grant 1191434.

 \vspace{1cm}
 \appendix

 \begin{boldmath}
 	\section{Representations and qualitative aspects} \label{APPENDIX:A}
 \end{boldmath}
 In this appendix we collected our findings of analytical estimates of $N_{1}^{I}(z)$ and $N_{1}^{II}(z,\dY)$. We believe that these estimates   show that these functions are small  and illustrate  our main result: the practical approach to solution of the BK non-linear equation in which we can find analytically the main part of the solution leaving the small corrections to be treated numerically.

~
~

\begin{boldmath}
	\subsubsection{$N_{1}^{I}(z)$ in the geometrical scaling region}
\end{boldmath}
Introducing  $
	N_{1}^{I}(z)\,=\, C_{1}^{I}\, (N^{I}_{0}(z)-1)\, +\, \Delta_{1}^{I}(z)$, with
\begin{equation*}\label{CONTRI:O1:R1}
	\Delta_{1}^{I}(z)\, =\, \int_{\epsilon-i\infty}^{\epsilon+i\infty}\dfrac{d\ga}{2\pi i}e^{(z+\zeta-z_{0})\ga}\underbrace{m(0)e^{\frac{1}{2}\kappa \ga^{2}}\int_{0}^{\ga}d\ga'\dfrac{a_{0}(\ga')}{m_{0}(\ga')}}_{g_{1}(\ga)}
\end{equation*}
and with $N_{0}^{I}(z)-1=C_{1}\, \Delta_{01}(z)$  where  $C_{1}=-4m(0)/\sqrt{2\pi\kappa}$ and   $\Delta_{01}(z)=e^{-\frac{(z+\zeta-z_{0})^{2}}{2\kappa}}$  is asymptotic formula of Ref.\cite{LT},  we have	$N_{1}^{I}(z)\,=\, \widetilde{C}_{1}\, \Delta_{01}(z)\, +\, \Delta_{1}^{I}(z)$
where $\widetilde{C}_{1}=- (\Delta_{1}^{I}(z^{A}_{0})/\Delta_{01}(z^{A}_{0}))$, with $z_0^{A}=\xi_{0}^{A}$.

 From the equation for $g_{1}(\ga)$:
\begin{equation*}
	\dfrac{dg_{1}}{d\ga}\, =\, \kappa \ga g_{1}\, +\, e^{2\psi(1)\ga}\dfrac{\Gamma(1-\ga)}{\Gamma(1+\ga)}a_{0}^{I}(\ga)
\end{equation*}
we can conclude
\begin{equation*}\label{REL1:SOL1}
	\kappa\dfrac{d\Delta_{1}^{I}}{dz}\, +\, (z+\zeta-z_{0})\Delta_{1}^{I}\, =\,-\, \underbrace{ \int_{\epsilon-i\infty}^{\epsilon+i\infty}\dfrac{d\ga}{2\pi i}\, u^{-\ga}e^{2\psi(1)\ga}\, a_{0}^{I}(\ga)\dfrac{\Gamma(1-\ga)}{\Gamma(1+\ga)}}_{H^{I}_{1}(z)}
\end{equation*}
with $u=e^{-(z+\zeta-z_{0})}$. The solution to this equation has the form:
\begin{equation*}\label{EXPRE:SOL1}
	\Delta_{1}^{I}(z)\, =\,  -\dfrac{1}{\kappa}\,\Delta_{01}(z) \int_{z_{0}^{A}}^{z}\dfrac{1}{\Delta_{01}(z')}H_{1}^{I}(z')\, dz'\, +\,  (\Delta_{1}^{I}(z^{A}_{0})/\Delta_{01}(z^{A}_{0}))  
\end{equation*}
which leads us
\begin{equation}\label{SOL1:FINAL:A}
	N_{1}^{I}(z)\, =\,   -\left(\dfrac{1}{\kappa} \int_{z_{0}^{A}}^{z}\dfrac{1}{\Delta_{01}(z')}H_{1}^{I}(z')\, dz'\, \right) \,\Delta_{01}(z)
\end{equation}
It is worthwhile mentioning that 
$N_{1}^{I}(z)$ satisfies
\begin{equation}\label{ODE:CONTRI:SUP} \dfrac{dN_{1}^{I}}{dz}+\dfrac{(z+\zeta-z_{0})}{\kappa}\, N_{1}^{I}\, =\, - \dfrac{1}{\kappa}H_{1}^{I}(z),
\end{equation}
with leads to a general case of 
 \eq{ODE:GEN:R1}).
  Note that $\Delta_{10}(z)$ is the homogeneous solution  of \eq{ODE:CONTRI:SUP}.

On the other hands, the non-homogeneous term $H_{1}^{I}(z)$ can be expressed in term of convolution product between Mellin transforms (see  formula {\bf 6.1.14} of Ref.\cite{BATEMAN}) as follows:
\begin{equation}\label{MELLIN:CONVOLUCION}
	H_{1}^{I}(z)\, =\,  e^{\psi(1)}e^{(z+\zeta-z_{0})/2}\, \int_{0}^{\infty}B_{0}^{I}(\eta)\, J_{1}\left(2\,  e^{\psi(1)}e^{(z+\zeta-z_{0})/2}\sqrt{\eta} \right)\, \dfrac{d\eta}{\sqrt{\eta}}
\end{equation}
with $B_{0}^{I}(u)=\frac{dA_{0}^{I}}{d\zz}$. Form \eq{M1I11} we have $B_{0}^{I}(u)=2\, M_{0}^{I}(\zz)\, \int_{0}^{\zz}d\zz''\, M_{0}^{I}(\zz'')$, where
\begin{equation}\label{EX:INT:SOL1}
\int_{0}^{\zz}d\zz''\, M_{0}^{I}(\zz'')\, =\, \int_{\epsilon-i\infty}^{\epsilon+i\infty}\dfrac{d\ga}{2\pi i}e^{(\zz+\zeta)\ga}\dfrac{1}{\ga}m_{0}^{I}(\ga)\, -\, \int_{\epsilon-i\infty}^{\epsilon+i\infty}\dfrac{d\ga}{2\pi i}e^{\zeta\ga}\dfrac{1}{\ga}m_{0}^{I}(\ga)
\end{equation}
Assuming that the principal contribution is given by the constant term in  \eq{EX:INT:SOL1}, we estimate  \eq{MELLIN:CONVOLUCION} considering the following approximation
\begin{equation*}
	\widetilde{H}_{1}^{I}(z)\, =\, c_{1}^{I}(\zeta)\, e^{\psi(1)}e^{(z+\zeta-z_{0})/2}\, \int_{0}^{\infty}M_{0}^{I}(\zz') J_{1}\left(2\,  e^{\psi(1)}e^{(z+\zeta-z_{0})/2}\sqrt{\eta} \right)\, \dfrac{d\eta}{\sqrt{\eta}}
\end{equation*}
with $c_{1}^{I}(\zeta)= -\, 2\, \int_{\epsilon-i\infty}^{\epsilon+i\infty}\frac{d\ga}{2\pi i}e^{\zeta\ga}\frac{1}{\ga}m_{0}^{I}(\ga)$, and $M^{I}_{0}(\zz')\, =\, \int_{\epsilon-i\infty}^{\epsilon+i\infty}\frac{d\ga}{2\pi i}\eta^{-\ga}m_{0}^{I}(\ga)$. Introducing $r'= e^{(z+\zeta)/2}$ and $\eta'=\sqrt{\eta}$, we  can rewrite 
\begin{equation}\label{APROX:NHOMO:S1}
	\widetilde{H}_{1}^{I}(z)\, =\, c_{1}^{I}(\zeta)\, e^{(z+\zeta)/2}\, \int_{0}^{\infty}M_{0}^{I}(\zz'')\, J_{1}\left(r'\, \eta' \right)\, d\eta'
\end{equation}
with $M_{1}^{I}(\zz'')\, =\, \int_{\epsilon-i\infty}^{\epsilon+i\infty}\frac{d\ga}{2\pi i}(\eta')^{-2\ga}m_{0}^{I}(\ga)$. Taking $\ga'=2\ga$ and changing  the order of  integration, \eq{APROX:NHOMO:S1} becomes
\begin{equation*}
	\widetilde{H}_{1}^{I}(z)\, =\, \frac{1}{2}c_{1}^{I}(\zeta)\, e^{(z+\zeta)/2}\, \int_{\epsilon'-i\infty}^{\epsilon'+i\infty}\dfrac{d\ga'}{2\pi i} \left(\int_{0}^{\infty}(\eta')^{-\ga'}\, J_{1}\left(r'\, \eta' \right)\, d\eta'\right)\, m_{0}^{I}\left( \frac{\ga}{2} \right)
\end{equation*}
From formula {\bf 6.561.14} of Ref.\cite{RY} follows
\begin{equation*}
	\widetilde{H}_{1}^{I}(z)\, =\, \frac{m(0)}{4}c_{1}^{I}(\zeta)\,  \int_{\epsilon'-i\infty}^{\epsilon'+i\infty}\dfrac{d\ga'}{2\pi i} \left(\frac{r'\, e^{-\psi(1)}}{2} \right)^{\ga'}\, e^{\frac{\kappa}{8}(\ga')^{2}}
\end{equation*}
with $r'e^{-\psi(1)}/2\, =\, e^{(z+\zeta-z_{0})/2}$, and therefore (use { \bf 7.2.1}  of Ref.\cite{BATEMAN})  we have
\begin{equation}\label{APROX:NHOMO:S1:B}
	\widetilde{H}_{1}^{I}(z)\, =\, \frac{m(0)}{2\sqrt{2\pi \kappa}}c_{1}^{I}(\zeta)\, \Delta_{01}(z).
\end{equation}
Plugging this approximation into \eq{SOL1:FINAL:A} we obtain the following estimates:
\begin{equation}\label{ESTI:SOL1:APEN}
		N_{1}^{I}(z)\, =\,   \left(\dfrac{c_{1}^{I}(\zeta)\, m(0)}{2\kappa\sqrt{2\pi\kappa}}\, (z-z_{0}^{A})\, +\, o(\vert z \vert) \right) \,\Delta_{01}(z)
\end{equation}
where the corrections, which were incorporated in $o(\vert z\vert)$, are obtained from the analysis of the contribution given by the term
\begin{equation*}
	m(0)\, M_{0}^{I}(z)\, \int_{\epsilon-i\infty}^{\epsilon+i\infty}\dfrac{d\ga}{2\pi i}\, e^{(z+\zeta-2\psi(1))\ga}\, e^{\frac{\kappa}{2}\ga^{2}}\dfrac{\Gamma(\ga)}{\Gamma(1-\ga)}\, =\, \dfrac{2m(0)}{\sqrt{2\pi\kappa}}\, M_{0}^{I}(z)\, \int_{0}^{\infty}e^{-\frac{1}{2\kappa}\ln^{2}(t')}J_{0}\left( 2\sqrt{\dfrac{u}{t}} \right)\, \dfrac{dt'}{t'}
\end{equation*}
with $u=e^{-(z+\zeta-z_{0})}$.  The detailed analysis of these contributions we will publish in a separate paper
in the nearest future. However, one can see that \eq{ESTI:SOL1:APEN} leads to a contribution, which is proportional to $\Delta_{01}(z)$. Therefore, it is small at least at large $z$.

 ~
 
 ~
 
\begin{boldmath}
	\subsubsection{$N_{1}^{II}(z,\dY)$   in the region II}\label{APPENDIX:B}
\end{boldmath}
Introducing  $h_{1}(\dY)\, =\, -\int_{0}^{\dY}dt\tilde{\Phi}_{1}^{II}(t)$ and $g(\ga,\dY)=2m(0)\, e^{\frac{1}{2}\kappa\ga^{2}}\int_{0}^{\dY}dt\frac{a_{0}^{II}(\ga+\dY-t,t)}{m_{0}(\ga+\dY-t)}$  we write $N_{1}^{II}(z,\dY)\, =\, h(\dY)(N_{0}^{II}(z,\dY)-1)\, +\, N_{1,p}(z,\dY)$ with  $N_{1,p}(z,\dY)=\int_{\epsilon-i\infty}^{\epsilon+i\infty}\frac{d\ga}{2\pi i}e^{(z+\zeta-z_{0})\ga}g(\ga,\dY)$, where $g(\ga,\dY)$  satisfies
 \begin{equation}\label{MELLIN:EQ:P1}
  \dfrac{\pp g}{\pp\dY}-\dfrac{\pp g}{\pp \ga}\, =\, -\kappa\, \ga\, g\, +\,  2e^{\psi(1)\ga}\dfrac{\Gamma(1-\ga)}{\Gamma(1+\ga)}\, a_{0}^{II}(\ga,\dY)
 \end{equation}
For  $H^{II}_{1}(z,\dY)\, =\, 2\, \int_{\epsilon-i\infty}^{\epsilon+i\infty}\frac{d\ga}{2\pi i}u^{-\ga}e^{2\psi(1)\ga}a_{0}^{II}(\ga,\dY)$ with  $u=e^{z+\zeta-z_{0}}$ \eq{MELLIN:EQ:P1} can be rewriiten as   the following partial differential equation:
 \begin{equation*}\label{EQ:P1:COORD}
 	\dfrac{\pp N_{1,p}}{\pp \dY}\, +\, (z+\zeta-z_{0})\, N_{1,p}\, +\, \kappa\, \dfrac{\pp N_{1,p}}{\pp z}\, =\,  H_{1}^{II}(z,\dY) 
 \end{equation*}
 or equivalently
  \begin{equation}\label{EQ:P1:COORDB}
 \dfrac{\pp \mu_{1}^{II}}{\pp \dY}\, +\, \kappa \dfrac{\pp \mu_{1}^{II}}{\pp z}\, =\, \dfrac{1}{ \Delta_{10}(z)} H_{1}^{II}(z,\dY) 
 \end{equation}
where $\mu_{1}^{II}(z,\dY)\, =\, N_{1,p}(z,\dY)/\Delta_{01}(z)$, with $\Delta_{01}(z)$ the  solution of Ref.\cite{LT}. From $N_{1}^{II}(z,\dY=0)=0$, the solution can be  written as follows
\begin{equation}\label{EXP:SOL:P1}
	N_{1}^{II}(z,\dY)\, =\, h_{1}(\dY)\, (N_{0}^{II}(z,\dY)-1)\, +\,  \Delta_{01}(z)\, \int_{0}^{\dY}\dfrac{H_{1}^{II}(z-\kappa \dY+\kappa t,t)}{\Delta_{01}(z-\kappa\dY+\kappa t)}\, dt
\end{equation}
We can express $H_{1}^{II}(z,t)$  similarly to \eq{MELLIN:CONVOLUCION}, as the following convolution product :
\begin{equation}\label{MELLIN:CONVOLUCION:R2}
	H_{1}^{II}(z,t)\, =\,  2\, e^{\psi(1)}e^{(z+\zeta-z_{0})/2}\, \int_{0}^{\infty}B_{0}^{II}(\eta\, ,\, t)\, J_{1}\left(2\,  e^{\psi(1)}e^{(z+\zeta-z_{0})/2}\sqrt{\eta} \right)\, \dfrac{d\eta}{\sqrt{\eta}}
\end{equation}
with $B_{0}^{II}(\eta,t)\, =\,  M_{0}^{II}(z',t)\, \int_{z'}^{\infty}d\zz'' M^{II}_{0}(\zz'',t)$ (see \eq{M1I23}). For $\epsilon<0$, we have
\begin{equation}\label{CONVO:PARTE2:APEN} 
B_{0}^{II}(\eta,t)\, =\, -\dfrac{1}{4}\, (f_{1}(t))^{2}\, \left( \int_{\epsilon-i\infty}^{\epsilon+i\infty}\dfrac{d\ga}{2\pi i}\eta^{-\ga}\dfrac{\Gamma(1+\ga)}{\Gamma(1-\ga)}\, \Gamma(-(\ga+t)) \right)\left( \int_{\epsilon-i\infty}^{\epsilon+i\infty}\dfrac{d\ga'}{2\pi i}\eta^{-\ga'}\dfrac{\Gamma(\ga')}{\Gamma(1-\ga')}\, \Gamma(-(\ga'+ t)) \right)
\end{equation}
with $\eta=e^{(z'-\kappa t)}$  (see the variable in the Mellin transform presented in \eq{MII4}).  Introducing $G(\eta,t)$ such that $B_{0}^{II}(\eta,t)\, =\, -\frac{1}{4}(f_{1}(t))^{2}\, G(\eta,t)$,   grouping and using \eq{MELLIN:CONVOLUCION:R2},  we can rewrite the second term of \eq{EXP:SOL:P1} in  the form:
\begin{equation}\label{EX:SOL:P1B}
	\begin{array}{l}
	\Delta_{01}(z)\, \displaystyle{\int_{0}^{\dY}\dfrac{H_{1}^{II}(z-\kappa \dY+\kappa t,t)}{\Delta_{01}(z-\kappa\dY+\kappa t)}\, dt}\, =\,- \frac{1}{4} \exp\left(  -\frac{1}{2}(z+\zeta-z_{0})\, \dY\right)\, \exp\left(-\frac{1}{2}(\xi+\zeta-z_{0})\dY\right)\times   \\
	  \,e^{(\xi + \zeta)/2}\displaystyle{\left(\int_{0}^{\dY}\, f_{1}(t)\left( \dfrac{r^{2}Q^{2}_{s}(Y_{A}, b)}{4}\right)^{t}\,e^{\kappa\,t/2}\left[\int_{0}^{\infty}G(\eta,t)\, J_{1}\left(r\, Q_{s}(Y_{A}, b)\, e^{\kappa\, t/2}\, \sqrt{\eta}  \right)\, \dfrac{d\eta}{\sqrt{\eta}} \right]\, dt\right)}	  	
	 \end{array}
\end{equation}

From \eq{EX:SOL:P1B} we see that the second term in \eq{EXP:SOL:P1} is proportional to $\Delta_{01}(z)$ and, therefore, it is not larger than the first term in this equation. However,  we expect from \eq{MII7} that the factor in parentheses is rather small, being of the order of  $ \exp\left(-\dfrac{r^{2}Q^2_{s}(Y_{A}, b)}{2}  \right)$.  Bearing this in mind we 
 assume that the contribution of \eq{EX:SOL:P1B}  is smaller than the first term in \eq{EXP:SOL:P1} and, therefore, we can trust \eq{NII10}.
We are planning to give a more complete proof of this smallness in our further publications.



\begin{thebibliography}{99}

\bibitem{GLR}
L. V. Gribov, E. M. Levin and M. G. Ryskin,
Phys. Rep. {\bf 100} (1983) 1.


\bibitem{MUQI}
A. H. Mueller and J. Qiu,
Nucl. Phys. {\bf B268} (1986) 427.

\bibitem{MV}
L. McLerran and R. Venugopalan,
Phys. Rev. {\bf D49} (1994) 2233, 3352; {\bf D50} (1994) 2225;
{\bf D53} (1996) 458;\\ {\bf D59} (1999) 09400.

\bibitem{MUCD}
 A.~H.~Mueller,
  Nucl.\ Phys.\  B {\bf 415}, 373 (1994);
  Nucl.\ Phys.\  B {\bf 437} (1995) 107
  [arXiv:hep-ph/9408245].
\bibitem{KLBOOK}
 Y.~V.~Kovchegov and E.~Levin,
  ``Quantum chromodynamics at high energy,''    Cambridge Monographs on Particle Physics, Nuclear Physics and Cosmology, Cambridge University Press, 2012 and references therein.
\bibitem{BK}
I.~Balitsky,
[arXiv:hep-ph/9509348];\,\,
{\it Phys.\ Rev.} {\bf D60}, 014020 (1999)
[arXiv:hep-ph/9812311];\,\,\,\,
Y.~V.~Kovchegov,
{\it Phys.\ Rev.}  {\bf D60}, 034008  (1999),
[arXiv:hep-ph/9901281].


\bibitem{JIMWLK}
 J.~Jalilian-Marian, A.~Kovner, A.~Leonidov and H.~Weigert,
  Nucl.\ Phys.\ B {\bf 504}, 415 (1997)
  [hep-ph/9701284];\,\,\,
J.~Jalilian-Marian, A.~Kovner, A.~Leonidov and H.~Weigert,
  Phys.\ Rev.\ D {\bf 59}, 014014 (1998)
  [hep-ph/9706377\,\,\,
J.~Jalilian-Marian, A.~Kovner and H.~Weigert,
  Phys.\ Rev.\ D {\bf 59}, 014015 (1998)
  [hep-ph/9709432]\,\,\,
 A.~Kovner, J.~G.~Milhano and H.~Weigert,
 {\it  Phys.\ Rev.}  {\bf D62}, 114005 (2000),
  [arXiv:hep-ph/0004014]\,; \,\,\,
E.~Iancu, A.~Leonidov and L.~D.~McLerran,
{\it  Phys.\ Lett.}\,  {\bf B510}, 133 (2001);
[arXiv:hep-ph/0102009];\,\, {\it  Nucl.\ Phys.}\,  {\bf A692}, 583
(2001),
[arXiv:hep-ph/0011241];\,\,\,
E.~Ferreiro, E.~Iancu, A.~Leonidov and L.~McLerran,
 {\it  Nucl.\ Phys.}\  {\bf A703}, 489 (2002),
  [arXiv:hep-ph/0109115];\,\,\,
H.~Weigert,
{\it  Nucl.\ Phys.}  {\bf A703}, 823 (2002),
[arXiv:hep-ph/0004044].
\bibitem{REV}
J.~L.~Albacete and C.~Marquet,
  Prog.\ Part.\ Nucl.\ Phys.\  {\bf 76} (2014) 1
  [arXiv:1401.4866 [hep-ph]].

\bibitem{LT}
  E.~Levin and K.~Tuchin,
  Nucl.\ Phys.\  B {\bf 573} (2000) 833
  [arXiv:hep-ph/9908317].
  \bibitem{BALE}
J.~Bartels, E.~Levin,
  Nucl.\ Phys.\  {\bf B387 } (1992)  617-637.
  \bibitem{SGBK}\,\,
 A.~M.~Stasto, K.~J.~Golec-Biernat, J.~Kwiecinski,
  Phys.\ Rev.\ Lett.\  {\bf 86 } (2001)  596-599,
  [hep-ph/0007192];  
  \bibitem{IIM}
 E.~Iancu, K.~Itakura, L.~McLerran,
  Nucl.\ Phys.\  {\bf A708 } (2002)  327-352.
  [hep-ph/0203137]  
\bibitem{MUT}
A.~H.~Mueller and D.~N.~Triantafyllopoulos,
{\it Nucl.\ Phys.} \, {\bf B640} (2002) 331
[arXiv:hep-ph/0205167];\,\,D.~N.~Triantafyllopoulos,
{\it Nucl.\ Phys.}\,  {\bf B648} (2003) 293
[arXiv:hep-ph/0209121].

 \bibitem{BFKL}
   V.~S. Fadin, E.~A. Kuraev and L.~N. Lipatov,
\newblock Phys. Lett. {\bf B60}, 50 (1975);\,\,\,
E.~A. Kuraev, L.~N. Lipatov and V.~S. Fadin,
\newblock Sov. Phys. JETP {\bf 45}, 199 (1977),
\newblock [Zh. Eksp. Teor. Fiz.72,377(1977)];\,\,\,
I.~I. Balitsky and L.~N. Lipatov,
\newblock Sov. J. Nucl. Phys. {\bf 28}, 822 (1978),
\newblock [Yad. Fiz.28,1597(1978)].
 \bibitem{LIP}
 L.~N.~Lipatov,
  Sov.\ Phys.\ JETP {\bf 63}, 904 (1986)
  [Zh.\ Eksp.\ Teor.\ Fiz.\  {\bf 90}, 1536 (1986)].


\bibitem{GOST}
K.~J.~Golec-Biernat and A.~M.~Stasto,
Nucl. Phys. B \textbf{668} (2003), 345-363
[arXiv:hep-ph/0306279 [hep-ph]].
\bibitem{BEST}
J.~Berger and A.~Stasto,
Phys. Rev. D \textbf{83} (2011), 034015
[arXiv:1010.0671 [hep-ph]].


\bibitem{KW1}
 A.~Kovner and U.~A.~Wiedemann,
  Phys.\ Rev.\ D {\bf 66}, 051502 (2002)
  [hep-ph/0112140].\,\,\,
  
  \bibitem{KW2}
 A.~Kovner and U.~A.~Wiedemann,  
  Phys.\ Rev.\ D {\bf 66}, 034031 (2002)
  [hep-ph/0204277];\,\,\.,
  \bibitem{KW3}
 A.~Kovner and U.~A.~Wiedemann,  
  Phys.\ Lett.\ B {\bf 551}, 311 (2003)
  [hep-ph/0207335].
  \bibitem{FIIM} 
  E.~Ferreiro, E.~Iancu, K.~Itakura and L.~McLerran,
  Nucl.\ Phys.\ A {\bf 710}, 373 (2002)
  [hep-ph/0206241].
  
  \bibitem{FROI}
M.~Froissart, 
 {\bf 123} (1961) 1053; \\
~A. ~Martin, {``Scattering Theory: Unitarity, Analitysity and Crossing."}
Lecture Notes in Physics, Springer-Verlag,  Berlin-Heidelberg-New-York,
1969. 
\bibitem{IIMU}
E.~Iancu, K.~Itakura and S.~Munier,
  Phys.\ Lett.\ B {\bf 590}, 199 (2004)
  [hep-ph/0310338].


\bibitem{BKL}
S.~Bondarenko, M.~Kozlov, E.~Levin,
  Nucl.\ Phys.\  {\bf A727 } (2003)  139-178,
  [hep-ph/0305150] .
\bibitem{RESH}
 A.~H.~Rezaeian and I.~Schmidt,
  Phys.\ Rev.\ D {\bf 88} (2013) 074016
  [arXiv:1307.0825 [hep-ph]].
  \bibitem{CLP}
  C.~Contreras, E.~Levin and I.~Potashnikova,
Nucl. Phys. A \textbf{948} (2016), 1-18
[arXiv:1508.02544 [hep-ph]].


\bibitem{HE1}
	J.H. He, 
	Comput. Methods Appl. Mech. Engrg. {\bf 173} (1999) 257.  
	
	\bibitem{HE2}
	J.H. He, 
	 Int. J. Nonlinear Mech. {\bf 35} (2000) 37.
\bibitem{SPS}
R.~Saikia, P.~Phukan and J.~K.~Sarma,
[arXiv:2204.10111 [hep-ph]].
\bibitem{GBW}
	K. Golec-Biernat and M. W$\ddot{u}$sthoff,
	 \emph{Phys. Rev. D} {\bf 59}, 014017 (1998).

\bibitem{LTHI}
 E.~Levin and K.~Tuchin,
  Nucl.\ Phys.\ A {\bf 693}, 787 (2001)
  [hep-ph/0101275].
\bibitem{KLT}
A.~Kormilitzin, E.~Levin and S.~Tapia,
Nucl. Phys. A \textbf{872}, 245-264 (2011)
doi:10.1016/j.nuclphysa.2011.09.021
[arXiv:1106.3268 [hep-ph]].
\bibitem{CLM}
C.~Contreras, E.~Levin and R.~Meneses,
JHEP \textbf{10}, 138 (2014)
doi:10.1007/JHEP10(2014)138
[arXiv:1406.1212 [hep-ph]].
\bibitem{KOV}
Yu, V. Kovchegov, ,
 Phys.\, Rev.\, {\bf D  61}  (2000) 074018.
 

\bibitem{BATEMAN}
Harry Bateman, {\it Tables of  integral transforms}, McGraw-Hill book company, inc. 1954.




\bibitem{MUPE}
  S.~Munier and R.~B.~Peschanski,
  Phys.\ Rev.\ D {\bf 69}, 034008 (2004)
  [hep-ph/0310357];\,\,\,Phys.\ Rev.\ Lett.\  {\bf 91}, 232001 (2003)
  [hep-ph/0309177].


\bibitem{RY}
I. Gradstein and I. Ryzhik, 
{\it "Tables of Series, Products, and Integrals"}, Fifth Edition, Academic Press, London, 1994.
\end{thebibliography}
     \end{document}